\documentclass[floatfix, superscriptaddress, preprintnumbers, twocolumn, amsmath, amssymb, aps, prl]{revtex4-1}

\usepackage{graphicx, xcolor}
\usepackage[caption=false]{subfig}
\usepackage[utf8]{inputenc}
\usepackage{soul}
\usepackage{times}
\usepackage{multirow}
\usepackage{booktabs}
\usepackage{hhline}
\usepackage{rotating}
\usepackage{complexity}
\usepackage{color}

\newcommand{\bra}[1]{\left\langle #1 \right|} 
\newcommand{\ket}[1]{\left| #1 \right\rangle} 
\newcommand{\braket}[1]{\left\langle #1 \right\rangle}

\begin{document}

\title{Experimental device-independent certified randomness generation with an instrumental causal structure}

\author{Iris Agresti}
\affiliation{Dipartimento di Fisica, Sapienza Universit\`{a} di Roma,
Piazzale Aldo Moro 5, I-00185 Roma, Italy}
\author{Davide Poderini}
\affiliation{Dipartimento di Fisica, Sapienza Universit\`{a} di Roma,
Piazzale Aldo Moro 5, I-00185 Roma, Italy}
\author{Leonardo Guerini}
\affiliation{International Center of Theoretical Physics - South American Institute for Fundamental Research, Instituto de F\'isica Te\'orica - UNESP, R. Dr. Bento T. Ferraz 271, 01140-070, S\~ao Paulo, Brazil}
\author{Michele Mancusi}
\affiliation{Dipartimento di Fisica, Sapienza Universit\`{a} di Roma,
Piazzale Aldo Moro 5, I-00185 Roma, Italy}
\author{Gonzalo Carvacho}
\affiliation{Dipartimento di Fisica, Sapienza Universit\`{a} di Roma,
Piazzale Aldo Moro 5, I-00185 Roma, Italy}
\author{Leandro Aolita }
\affiliation{International Center of Theoretical Physics - South American Institute for Fundamental Research, Instituto de F\'isica Te\'orica - UNESP, R. Dr. Bento T. Ferraz 271, 01140-070, S\~ao Paulo, Brazil}
\affiliation{Instituto de F\'{i}sica, Universidade Federal do Rio de Janeiro, Caixa Postal 68528, Rio de Janeiro, RJ 21941-972, Brazil}
\author{Daniel Cavalcanti}
\affiliation{ICFO - Institut de Ciencies Fotoniques, The Barcelona Institute of Science and Technology, E-08860 Castelldefels, Barcelona, Spain}
\author{Rafael Chaves}
\affiliation{International Institute of Physics, Federal University of Rio Grande do Norte, 59078-970, P. O. Box 1613, Natal, Brazil}
\affiliation{School of Science and Technology, Federal University of Rio Grande do Norte, 59078-970 Natal, Brazil}
\author{Fabio Sciarrino}
\affiliation{Dipartimento di Fisica, Sapienza Universit\`{a} di Roma,
Piazzale Aldo Moro 5, I-00185 Roma, Italy}

\begin{abstract} 
The intrinsic random nature of quantum physics offers novel tools for the generation of random numbers, a central challenge for a plethora of fields.
Bell non-local correlations obtained by measurements on entangled states allow for the generation of bit strings whose randomness is guaranteed in a device-independent manner, i.e. without assumptions on the measurement and state-generation devices.
Here, we generate this strong form of certified randomness on a new platform: the so-called instrumental scenario, which is central to the field of causal inference.
First, we theoretically show that certified random bits, private against general quantum adversaries, can be extracted exploiting device-independent quantum instrumental-inequality violations. To that end, we adapt techniques previously developed for the Bell scenario. Then, we experimentally implement the corresponding randomness-generation protocol using entangled photons and active feed-forward of information. Moreover, we show that, for low levels of noise, our protocol offers an advantage over the simplest Bell-nonlocality protocol based on the Clauser-Horn-Shimony-Holt inequality.
\end{abstract}

\maketitle

\section{Introduction}
The generation of random numbers has applications in a wide range of fields, from scientific research -- e.g. to simulate physical systems -- to military scopes -- e.g. for effective cryptographic protocols -- and every-day concerns -- like ensuring privacy and gambling. From a classical point of view, the concept of randomness is tightly bound to the incomplete knowledge of a system; indeed, classical randomness has a subjective and epistemological nature and is erased when the system is completely known \cite{WIQ}.
Hence, classical algorithms can only generate pseudo-random numbers \cite{pseudo_random}, whose unpredictability relies on the complexity of the device generating them.
Besides, the certification of randomness is an elusive task, since the available tests can only verify the absence of specific patterns, while others may go undetected but still be known to an adversary \cite{statistical_test}.

\begin{figure*}[tb!]
\includegraphics[scale=0.45]{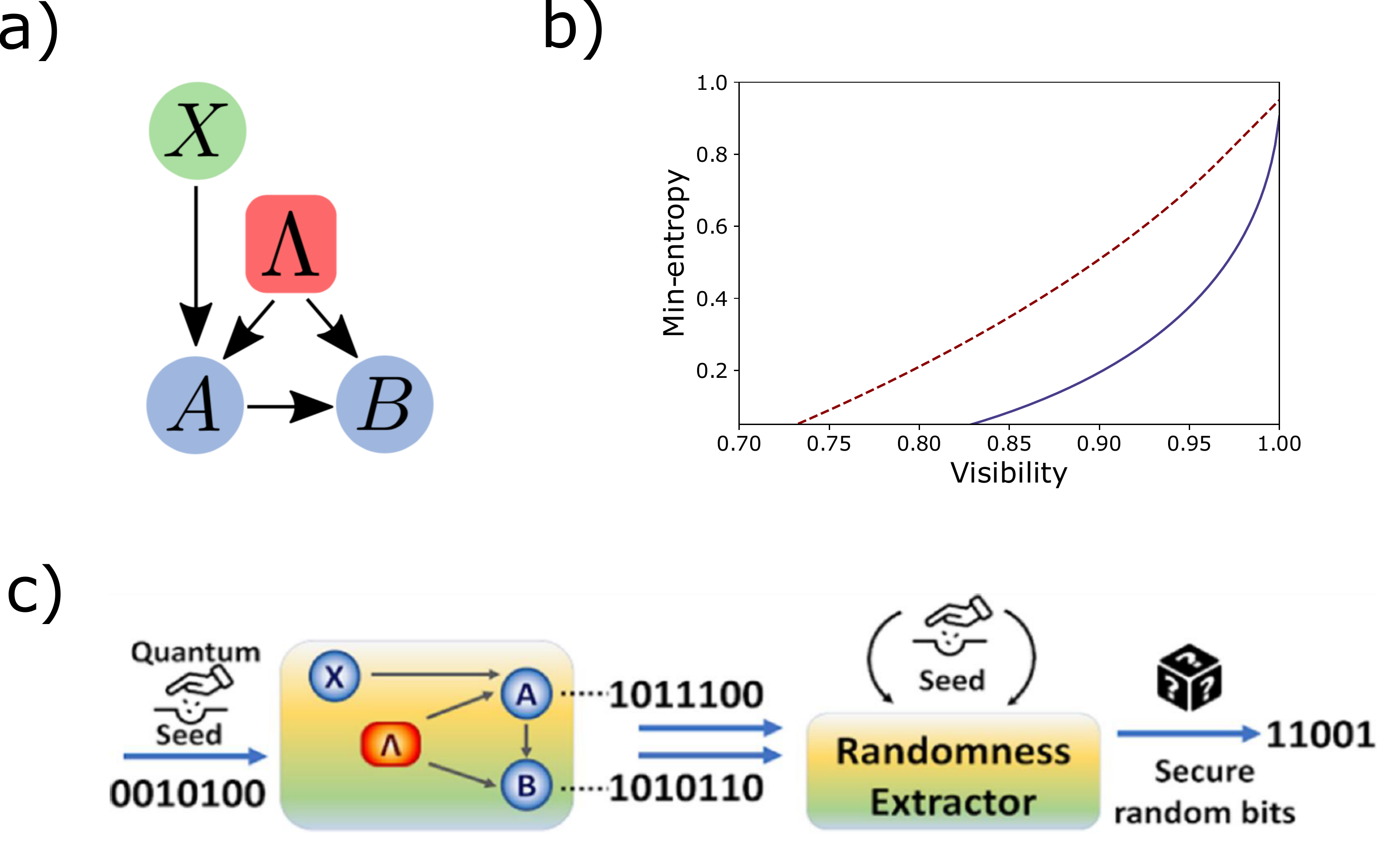}
\caption{\textbf{Randomness generation and certification protocol.} \textbf{a)} Instrumental causal structure represented as a directed acyclic graph \cite{pearlbook} (DAG), where each node represents a variable and the arrows link variables between which there is causal influence. In this case, X, A and B are observable, while $\Lambda$ is a latent variable. \textbf{b)} The plot shows the smooth min-entropy bound for the CHSH and the instrumental scenario (respectively dashed and continuous curve), in terms of the state visibility $v$, i.e. considering the extent of violation that would be given by the following state:  $\rho= v \ket{\psi^-}\bra{\psi^-} + (1-v) \frac{I}{4}$. The bounds were obtained through the analysis of \cite{friedman2019}, secure against general quantum adversaries, which was adapted to our case. The choice of parameters was the following: $n=10^{12}$, $\epsilon=\epsilon_{EA}=10^{-6}$, $\delta'=10^{-4}$ and $\gamma=1$. In detail, $n$ is the number of runs, $\epsilon$ is the smoothing parameter characterizing the min-entropy $\mathcal{H}^{\epsilon}_{min}$, $\epsilon_{EA}$ is the desired error probability of the entropy accumulation protocol, $\delta'$ is the experimental uncertainty on the evaluation of the violation $\mathcal{I}$ and $\gamma$ is the parameter of the Bernoulli distribution according to which we select \textit{test} and \textit{accumulation} runs throughout the protocol.  \textbf{c)} Simplified scheme of the proposed randomness generation and certification protocol (in the case $\gamma=1$): (i) initial seed generation (defining, at each run, Alice's choice among the operators), (ii) instrumental process implementation, (iii) classical randomness extractor. The initial seed is obtained from the random bits provided by the NIST Randomness Beacon \cite{beacon}. In the second stage, Alice's and Bob's outputs are collected and the corresponding value of the instrumental violation $\mathcal{I}^{*}$ is computed. If it is higher than a threshold set by the user, the smooth min-entropy is bounded by inequality (\ref{newbound}), otherwise the protocol aborts. The value of the min-entropy indicates the maximum number of certified random bits that can be extracted. At the end, if the protocol does not abort, the output strings are injected in a classical randomness extractor (Trevisan's extractor \cite{trevisan}) and the final string of certified random bits is obtained. The extractor's seed is as well provided by the NIST Randomness Beacon.}
\label{fig:concettuale}
\end{figure*}

On the other hand, randomness is intrinsic to quantum systems, which do not possess definite properties until these are measured. In real experiments, however, this intrinsic quantum randomness comes embedded with noise and lack of complete control over the device, compromising the security of a quantum random-number generator. A solution to that is to devise quantum protocols whose correctness can be certified in a device-independent (DI) manner, i.e. solely from the observed statistics and with no assumption whatsoever on the internal working of the experimental devices.
For instance, from the violation of a Bell inequality \cite{bell64, brunner14} one can ensure that the statistics of certain quantum experiments cannot be described in the classical terms of local deterministic models, hence being impossible to be deterministically predicted by any local observer.
Moreover, the extent of such violation can provide a lower bound on the certified randomness characterizing the measurement outputs of the two parties performing the Bell test, as shown in Ref. \cite{rng_certified}. 
Several other seminal works based on Bell inequalities have been developed \cite{Colbeck12, Gallego_2013, Brandao16, Ramanathan16, miller2016, vazirani2012, pan2018,vazirani2016QKD, chung2015, dupuis2016, friedman2019, christensen2013, friedman2019, continuous_rand, BSS14}, advancing the topics of \textit{randomness amplification} (the generation of near-perfect randomness from a weaker source), \textit{randomness expansion} (the expansion of a short initial random seed), and \textit{quantum key distribution} (sharing a common secret string through communication over public channels).
In particular, loophole-free Bell tests based on randomness generation protocol have been implemented \cite{christensen2013, bierhorst2018, rng_certified} and more advanced techniques have been developed to provide security against general adversarial attacks \cite{friedman2019, continuous_rand, Friedman2017}.

From a causal perspective, the non-classical behaviour revealed by a Bell test lies in the incompatibility of quantum predictions with our intuitive notion of cause and effect \cite{pearlbook,Wood2015,Chaves2015}. Given that the causal structure underlying a Bell-like scenario involves five variables (the measurement choices and outcomes for each of the two observers and a fifth variable representing the common cause of their correlations), it is natural to wonder whether a simpler causal structure could give rise to an analogous discrepancy between quantum and classical causal predictions \cite{henson2014, perspectivecausality}. 
The instrumental causal structure \cite{pearl1, Bonet}, where the two parties (A and B) are linked by a classical channel of communication (shown in Fig.\ref{fig:concettuale}-a), is the simplest model (in terms of the number of involved nodes) achieving this result \cite{instrumental}. This scenario has fundamental importance in causal inference, since it allows the estimation of causal influences even in the presence of unknown latent factors \cite{pearl1}.\\
\indent In this letter, we implement a device-independent random number generator based on instrumental correlations, secure against general quantum attacks \cite{friedman2019}.
Our protocol is device-independent in the sense that we do not make any assumption about the measurements and states used in the protocol, not even their dimension. We stress however that the implementation is assumed to respect the causal structure imposed by the instrumental scenario.
To implement the protocol in all of its parts, we have set up a classical extractor following the theoretical design by Trevisan \cite{trevisan}. Moreover, we prove that device-independent randomness generation protocols implemented in this scenario, for high state visibilities, can bring an advantage in the gain of random bits when compared to those based on on the simplest two-input-two-output Bell scenario, i.e. the Clauser-Horn-Shimony-Holt (CHSH) \cite{chsh_1969}.
Therefore, this work paves the way to further applications of the instrumental scenario in the field of device-independent protocols, which, until now, have relied primarily on Bell-like tests.

\section{Randomness certification via instrumental inequality violations}\label{sec2}

Let us first briefly review some previous results obtained in the context of Bell inequalities \cite{bell64}. In a CHSH scenario \cite{chsh_1969}, two parties, A and B, share a bipartite system and, without communicating to each other, perform local measurements on their subsystems. If A and B choose between two given operators each, i.e. ($A_{1}$, $A_{2}$) and ($B_{1}$, $B_{2}$) respectively, and then combine their data, the mean value of the operator $S=|\braket{A_{1},B_{1}}-\braket{A_{1},B_{2}}+\braket{A_{2},B_{1}}+\braket{A_{2},B_{2}}|$ should be upper-bounded by 2, for any deterministic model respecting a natural notion of locality. However, as proved in \cite{chsh_1969}, if A and B share an entangled state, they can get a value exceeding such bound, whose explanation requires the presence of non-classical correlations between the two parties. Hence, Bell inequalities have been adopted in \cite{rng_certified} to guarantee the intrinsic random nature of the measurements' outcomes, within a DI randomness generation and certification protocol.

In the instrumental causal model, which is depicted in Fig.\ref{fig:concettuale}-a, the two parties (Alice and Bob) still share a bipartite state.
Alice can choose among $l$ possible $d$-outcome measurements ($O_{A}^1$, ..., $O_{A}^l$), according to the \textit{instrument} variable $x$, which is independent from the shared correlations between Alice and Bob ($\Lambda$) and can assume $l$ different values, while Bob's choice among $d$ observables ($O_{B}^1$, ..., $O_{B}^d$) depends on Alice's outcome. 
In other words, as opposed to the spatially-separated correlations in a Bell-like scenario, the instrumental process constitutes a temporal scenario, with one-way communication of Alice's outcomes to select Bob's measurement.
Hence, due to the communication of Alice's outcome $a$ to Bob, Bob's outcome $b$ is not independent of $x$; however, the instrumental network specifies this influence to be indirect, formalized by the constraint $p(b|x,a,\lambda) = p(b|a,\lambda)$ and justifying the absence of an arrow from $X$ to $B$ in Fig. \ref{fig:concettuale}-a.

Similarly to Bell-like scenarios, the causal structure underlying an instrumental process imposes some constraints on the classical joint probabilities $\{p(a,b|x)\}_{a,b,x}$ that are compatible with it \cite{pearl1,Bonet} (the so-called instrumental inequalities).
In the particular case where the instrument $x$ can assume three different values (1,2,3), while $a$ and $b$ are dichotomic, the following inequality holds \cite{Bonet}: 
\begin{equation}
  \mathcal{I}=\braket{A}_1-\braket{B}_1+2\braket{B}_2-\braket{AB}_1+2\braket{AB}_3 \leq 3
  \label{instrumental322}
  \end{equation}
where $\braket{AB}_x= \sum_{a,b=0,1} (-1)^{a+b} p(a,b \vert x)$.
Remarkably, this inequality can be violated with the correlations produced by quantum instrumental causal models \cite{instrumental}, up to the maximal value of $\mathcal{I}=1+2\sqrt{2}$. Recently, the relationship of the instrumental processes with the Bell scenario has been  studied in \cite{himbeeck2018}.

\indent In this context, we rely on the fact that if a given set of statistics $\{p(a,b\vert x)\}_{a,b,x}$ violates inequality \eqref{instrumental322}, then the system shared by the two parties exhibits non-classical correlations that impose non-trivial constraints on the information a potential eavesdropper could obtain, represented in the probability distributions $\{p(a,b,e\vert x)\}_{a,b,e,x}$, where $e$ is the eventual side information of the eavesdropper.
Consequently, this restricts the values of the conditional min-entropy, a randomness quantifier defined as $\mathcal{H}_{min}=-\log_{2}[\sum_e P(e)\max_{a,b} P(a,b\vert e, x)]$ \cite{PM13}. Indeed, it is possible to obtain a lower-bound on the min-entropy, for each $x$, as a function $f_x$ of $\mathcal{I}$: $\mathcal{H}_{min}\geq f_x(\mathcal{I})$. For each $x$ and $\mathcal{I}$, the lower bound $f_x(\mathcal{I})$ can be computed via semidefinite programming (SDP), by applying to the instrumental case the numerical techniques developed in \cite{bound, rng_certified}. Such method was originally conceived for the Bell scenario (see Supplementary Information), but it can actually be applied to any casual model involving a shared bipartite system, on whose subsystems local measurements are performed.
The functions $f_x$ are convex and grow monotonically with $\mathcal{I}$; so, the higher the violation of inequality \eqref{instrumental322} is, the higher the min-entropy lower bound will be.
Nevertheless, in real experimental conditions, in order to evaluate the quantum violation extent $\mathcal{I}^{*}$ (or, analogously, the probability distribution $p^{*}(a,b|x)$) to compute $f_{x}$, several experimental runs are necessary.
Therefore, unless one makes the \textit{iid assumption} (i.e. all the experimental runs are assumed to be independent and neither the state source nor Alice's and Bob's measurement devices exhibit time-dependent behaviours), this bound $f_{x}(\mathcal{I^{*}})$ will not hold in the presence of an adversary that could include a (quantum) memory in the devices, introducing interdependences among the runs. Several device-independent protocols have been proposed so far addressing the most general non-\emph{iid} case \cite{reichardt2013, vazirani2014, miller2014}, but at the cost of a low feasibility.
Only very recently a feasible solution for the CHSH scenario has been proposed and used \cite{friedman2019, continuous_rand, Friedman2017}, resorting to the \textit{Entropy Accumulation Theorem (EAT)}, in order to deal with processes not necessarily made of independent and identically distributed runs.

\begin{figure*}[tb!]
\includegraphics[scale= 0.5]{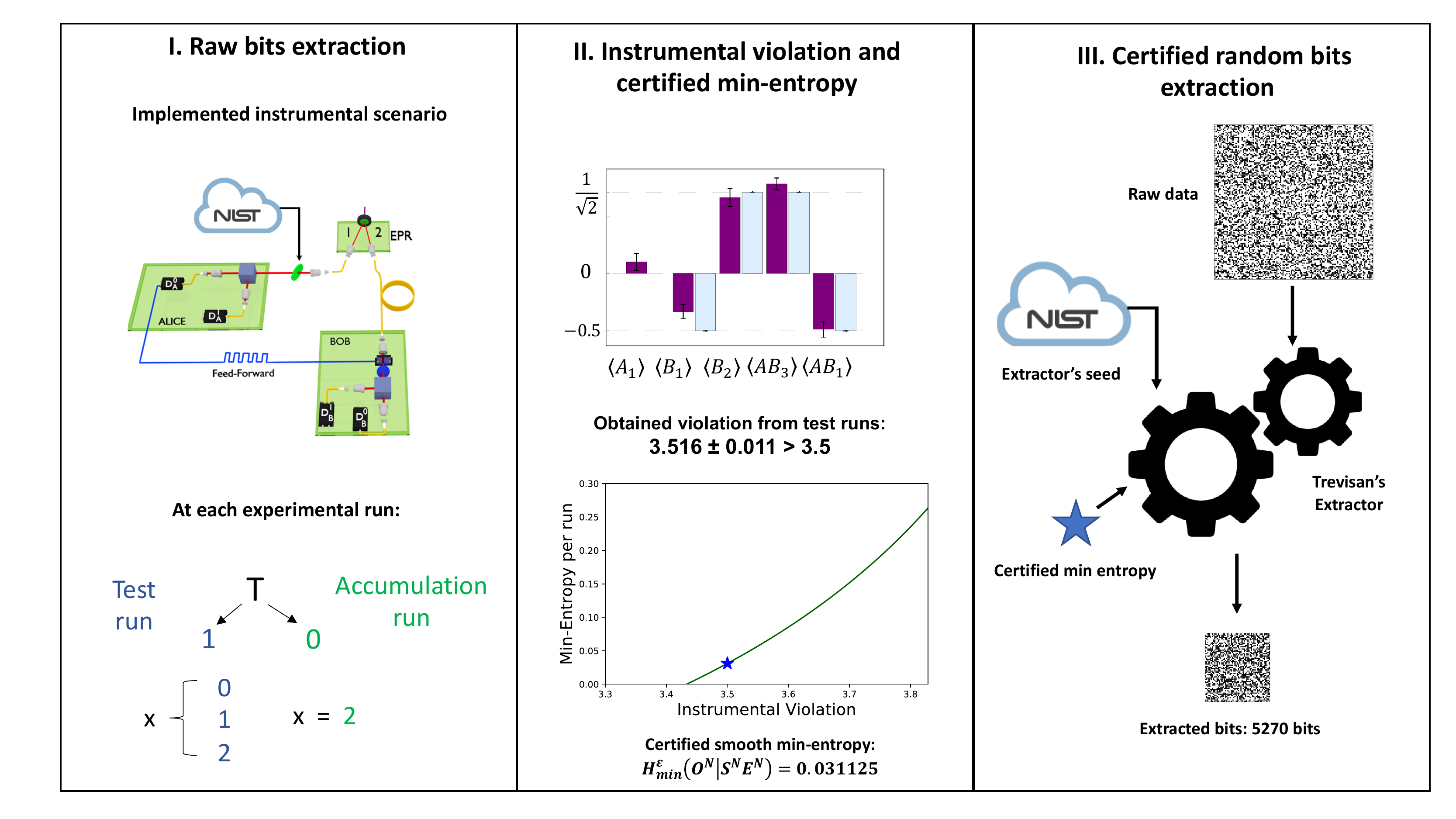}
\caption{\textbf{Implementation of the device-independent randomness certification protocol:} The implementation of our proposed protocol involves three steps. First of all, an instrumental process is implemented on a photonic platform. Then, for each round of the experiment, a binary random variable $T$ is evaluated. Specifically, $T$ can get value 1 with probability $\gamma$, previously chosen by the user (in our implementation, $\gamma=1$).
If $T=0$, the run is an \textit{accumulation} one, and $x$ is deterministically equal to 2. If $T=1$, the run is a \textit{test} run and $x$ is randomly chosen among 1, 2 and 3.  Note that, in our case, we only have \textit{test} runs. Secondly, after $n$ runs, through the bits collected in the test runs, we evaluate the corresponding instrumental violation and see whether it is higher than the expected violation for an honest implementation of the protocol, i.e. in a scenario with no eavesdroppers. In our case, we set the threshold to 3.5. If it is lower, the protocol aborts, otherwise, the protocol reaches the third stage, where we employ the Trevisan extractor, to extract the final certified random bit string. The extractor takes, as input, the raw data (weak randomness source), a random seed (given by the NIST Randomness Beacon \cite{beacon}) and the min-entropy of the input string. In the end, according to the classical extractor statistical error ($\epsilon_{ext}$) set by the user (in our case $10^{-6}$), the algorithm extracts $m$ truly random bits, with $m < n$.}
\label{fig:estrazione}
\end{figure*}

Here we adapt the technique developed in \cite{friedman2019} to the instrumental scenario, making our randomness certification secure against general quantum attacks.
According to the EAT, for a category of processes that comprehends also an implemented instrumental process composed of several runs, the following bound on the smooth min-entropy holds:
\begin{equation}
    \mathcal{H}_{min}^{\epsilon}(O^{n}|S^n E^n) > nt-\nu
    \sqrt{n},
    \label{newbound}
\end{equation}
where $O$ are the quantum systems given to the honest parties Alice and Bob, at each run, $S$ constitutes the leaked side-information, while $E$ represents any additional side information correlated to the initial state. Then, $t$ is a convex function which depends on the extent of the violation, expected by an honest, although noisy, implementation of the protocol, i.e. in a scenario with no eavesdropper ($I_{exp}$). On the other hand, $\nu$ depends also on the smoothing parameter $\epsilon$, which characterizes the smooth min-entropy $\mathcal{H}_{min}^{\epsilon}$, and $\epsilon_{EA}$, i.e. the desired error probability of the entropy accumulation protocol; in other words, either the protocol aborts with probability higher than $1-\epsilon_{EA}$ or the bound (\ref{newbound}) holds (for further detail, see the Supplementary Information).

\begin{figure}[h!]
\includegraphics[width=\columnwidth]{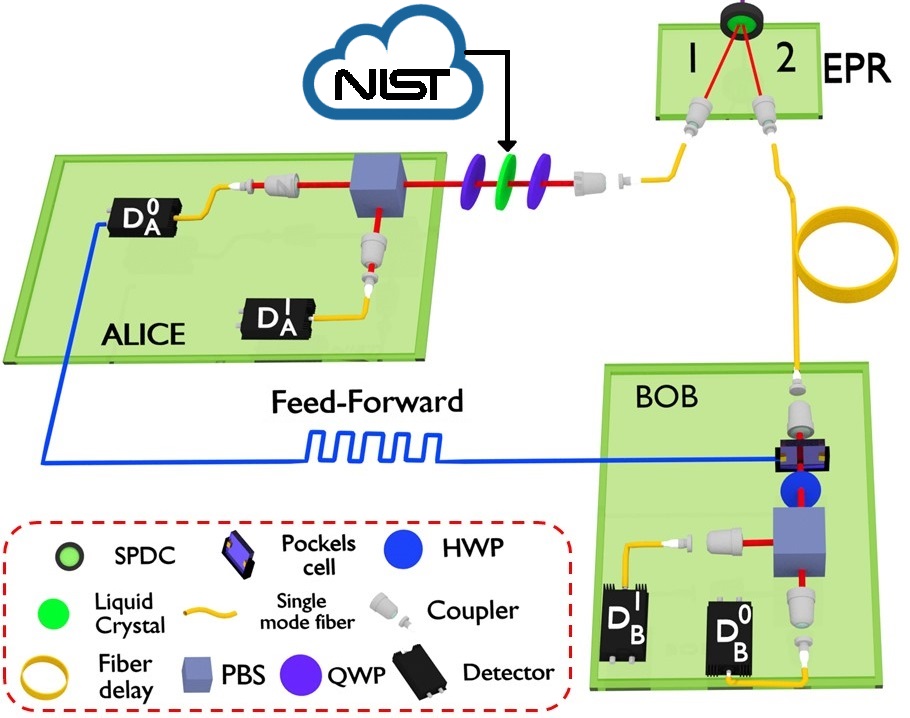}
\caption{ \textbf{Experimental apparatus}: A polarization-entangled photon pair is generated via spontaneous parametric down-conversion (SPDC) process in a nonlinear crystal. Photon $1$ is sent to Alice's station, where one of three observables ($O^{1}_{A}$, $O^{2}_{A}$ and $O^{3}_{A}$) is measured through a liquid crystal followed by a polarizing beam splitter (PBS). The choice of the observable relies on the random bits generated by the NIST Randomness Beacon \cite{beacon}. Detector $D_{A}^{0}$ acts as trigger for the application of a 1280 V voltage on the Pockels cell, whenever the measurement output $0$ is registered. The photon $2$ is delayed 600 ns before arriving to Bob's station by employing a single-mode fiber 125 m long. After leaving the fiber, the photon passes through the Pockels cell, followed by a fixed HWP at $56.25^{\circ}$ and a PBS. If the Pockels cell has been triggered (in case of A measurement outcome is $0$), its action combined to the fixed HWP in Bob's station allows us to perform $O^{1}_{B}$. Otherwise (if $A$ measurement outcome is $1$), the Pockels cell acts as the identity and we implement $O^{2}_{B}$.}
\label{fig:SETUP}
\end{figure}
Our protocol is implemented as follows (see Fig.\ref{fig:estrazione}): for each run, a random binary variable $T$ is drawn according to a Bernoulli distribution of parameter $\gamma$ (set by the user); if $T=0$, the run is an \textit{accumulation} run, so $x$ is deterministically set to 2 (which guarantees a higher $f(\mathcal{I})$, see the Supplementary Information); on the other hand, if $T=1$, the run is a \textit{test} run, so $x$ is randomly chosen among 1, 2 and 3. After $m$ test runs (with $m$ chosen by the user), the quantum instrumental violation is evaluated from the bits collected throughout the test runs and, if 
lower than $I_{exp}-\delta'$, ($\delta'$ being the experimental uncertainty on $\mathcal{I}$) the protocol aborts; otherwise the certified smooth min-entropy is bounded by inequality (\ref{newbound}).
This lower bound on the certified min entropy represents the maximum certified amount of bits that we can extract from our collected data. Hence, feeding the raw bit string and the $\mathcal{H}_{min}^{\epsilon}$ to the classical extractor \cite{trevisan}, the algorithm will output at most $\mathcal{H}_{min}^{\epsilon}(O^n|S^nE^n)$ certified random bits, the exact value depending on its internal error parameter $\epsilon_{ext}$. Specifically, we resorted to the classical extractor devised by Trevisan \cite{trevisan}.
This algorithm takes as inputs a \textit{weak randomness source}, in our case the $2n$ raw bits long string, and a seed, which is poly-logarithmic in the input size (our code for the classical extractor can be found at \cite{code} and, for a detailed description of the classical randomness extractor, see Supplementary Information).

\section{Experimental implementation of the protocol}

The device-independent random numbers generator, in our proposal, is made up of three main parts, which are seen as black boxes to the user: the state generation and Alice's and Bob's measurement stations. The causal correlations among these three stages are those of an instrumental scenario (see Fig. \ref{fig:concettuale}-a-c) and are implemented through the photonic platform depicted in Fig.\ref{fig:SETUP}.

Within this experimental apparatus, the qubits are encoded in the photon's polarization: horizontal ($\ket{H}$) and vertical ($\ket{V}$) polarizations represent, respectively, qubits $\ket{0}$ and $\ket{1}$, eigenstates of the Pauli matrix $\sigma_z$. 
A spontaneous parametric down-conversion (SPDC) process generates the two-photon maximally entangled state $\ket{\Psi^{-}}=\frac{\ket{HV}-\ket{VH}}{\sqrt{2}}$. 
One photon is sent to path 1, towards Alice's station, where an observable among $O_{A}^1$, $O_{A}^2$ and $O_{A}^3$ is measured, applying the proper voltage to a liquid crystal (LCD). 
The voltage must be chosen according to a random seed, made of a string of trits (indeed, in our case, we take $\gamma=1$, so $x$ is chosen among (1,2,3) at every run). 
This seed is obtained from the NIST Randomness Beacon \cite{beacon}, which provides 512 random bits per minute.
After Alice has performed her measurement, whenever she gets output 1 (i.e. $D_{A}^{0}$ registers an event), the detector's signal is split to reach the coincidence counter and, at the same time, trigger the Pockels cell on path 2. Bob's station is made of a Half-Wave Plate (HWP) followed by this fast electro-optical device. When no voltage is applied to the Pockels cell, Bob's operator is $O_{B}^1$ and, when it is turned on, there is a swift change to $O_{B}^2$ (the cell's time response is of the order of nanoseconds). In order to have the time to register Alice's output and select Bob's operator accordingly, the photon on path 2 is delayed, through a 125 meters long single-mode fiber.

The four detectors are synchronized in order to distinguish the coincidence counts generated by the entangled photons' pairs from the accidental counts. Let us note that our experimental implementation requires the \textit{fair sampling} assumption, due to our overall low detection efficiency. The measurement operators achieving maximal violation of $\mathcal{I}= 1+2\sqrt{2}$, when applied to the state $\ket{\psi^{-}}$, are the following: $O_{A}^{1}=-(\sigma_{z}-\sigma_{x})/\sqrt{2}$, $O_{A}^{2}=-\sigma_{x}$, $O_{A}^{3}=\sigma_{z}$ and $O_{B}^{1}=(\sigma_{x}-\sigma_{z})/\sqrt{2}$, $O_{B}^{2}=-(\sigma_{x}+\sigma_{z})/\sqrt{2}$.
After having implemented the instrumental scenario,  performed several experimental runs and collected the raw bits, we verified that the violation was higher than the chosen threshold $I_{exp}=3.5$ and evaluated the smooth min-entropy bound according to inequality (\ref{newbound}). In the end, we executed the classical randomness extractor devised by Trevisan \cite{trevisan}. The complete procedure is summarized in Fig.\ref{fig:estrazione}.

\begin{figure}[h!]
\includegraphics[width=\columnwidth]{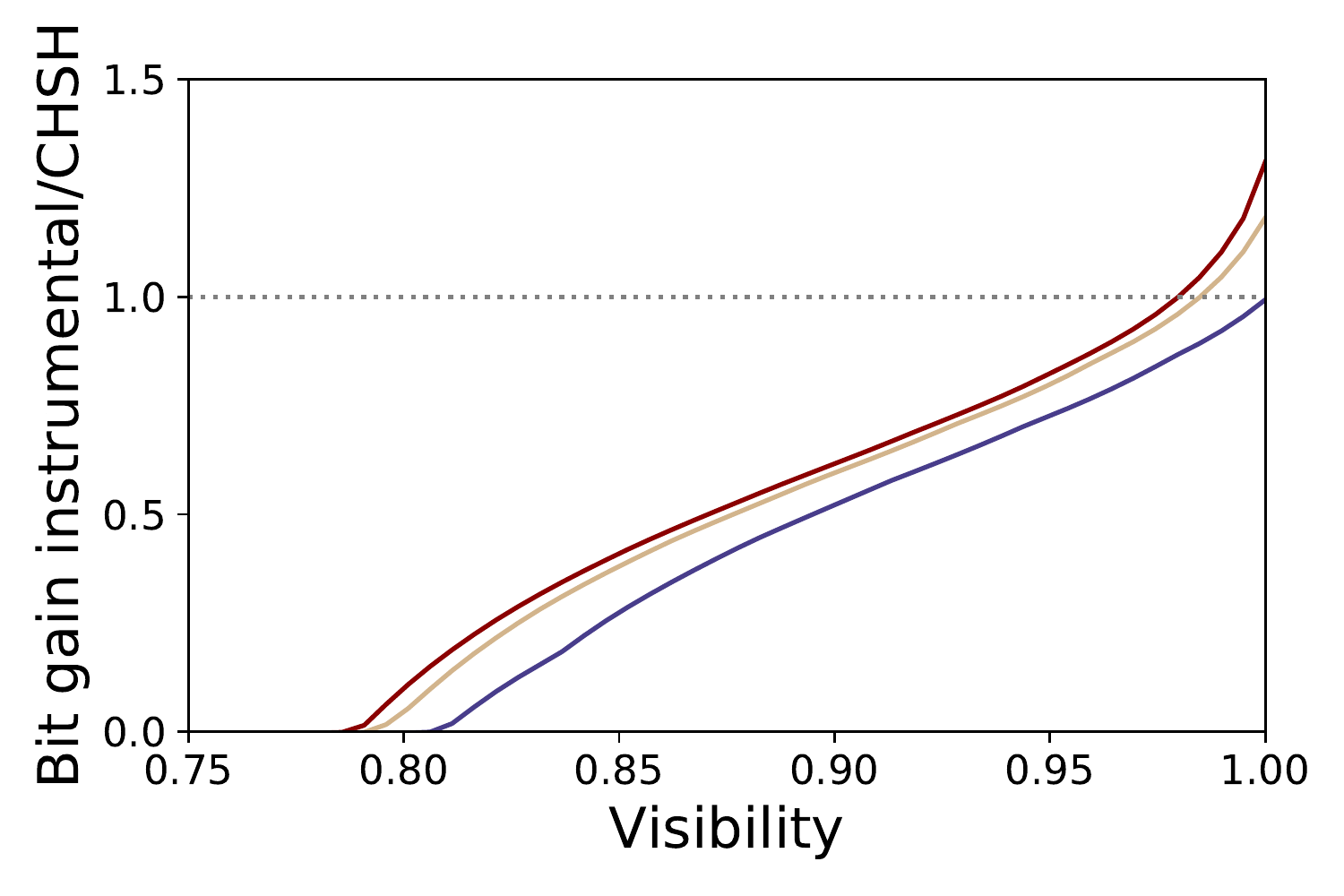}
\caption{\textbf{Comparison between CHSH and Instrumental random bits gain.} The plot shows the amount of random bits gained through the proposed protocol involving the instrumental scenario, over those gained in its regular CHSH based counterpart \cite{friedman2019, continuous_rand}, employing the same randomness source, which provides a given amount of initial bits to feed both the protocols. The different curves represent different amounts of invested random bits, in particular $n=10^{8}$ (blue, lowest curve), $n=10^{9}$ (golden, middle curve) and $n=10^{10}$ (red, highest curve), in terms of the state visibility $v$, i.e. corresponding to the extent of violation that would be given by the following state: $\rho= v \ket{\psi^-}\bra{\psi^-} + (1-v) \frac{I}{4}$. For each \textit{test run} of our protocol, we consider that the amount of invested bits to extract Alice's input $x$ is $log_{2}(3)$. These curves were obtained optimizing $\gamma$ for both scenarios.}
\label{fig:gain}
\end{figure}

\section{Results}

\subsection{Theoretical Results}
The DI random number generation protocol we propose for the instrumental scenario was developed adapting the pre-existing techniques for the Bell scenario \cite{friedman2019, continuous_rand}, and is secure against general quantum adversaries.
The most striking aspect of our protocol is that, as shown in Fig.\ref{fig:gain},  although in terms of min-entropy the CHSH scenario guarantees a higher bound (as it can be seen in Fig.\ref{fig:concettuale}-b), if we look at the randomness gain, considering an arbitrary randomness source which provides a given quantity of bits to use as input both of our protocol, as well as of its CHSH based counterpart, the instrumental scenario proves to be more convenient. In other words, more certified random bits can be extracted, investing the same amount of initial randomness. This happens in the regime of high state visibilities $v$ considering the following state $\rho= v \ket{\psi^-}\bra{\psi^-} + (1-v) \frac{I}{4}$ and large amounts of invested random bits, for example at least $10^9$ bits (golden curve), with a visibility of $\sim 0.98$.

\subsection{Experimental Results}

We implemented the instrumental scenario on a photonic platform and adopted the proposed quantum adversary-proof protocol in our experimental conditions. Specifically, we had 172095 experimental runs and set $\epsilon_{EA}=\epsilon=10^{-1}$ and the threshold to $I_{exp}=3.5$, with $\delta'=0.011$. Since the registered violation was of $3.516 \pm 0.011$, our smooth min-entropy bound, according to inequality (\ref{newbound}), was 0.031125, which  allowed us to gain, through the classical extractor, an overall number of 5270 random bits, with an error on the classical extractor of $\epsilon_{ext}=10^{-6}$. Note that each experimental run lasted $\sim 1s$ and the bottleneck of our implementation is the time response of the liquid crystal, $\sim 700$ ms, that implements Alice's operator. Hence, in principle, significantly higher rates can be reached on the same platform, adopting a fast electro-optical device also for Alice's station, with a response time of $\sim 100$ ns.

The length of the seed required by the classical extractor, as mentioned, is poly-logarithmic in the input size and its length also depends on the chosen error parameter $\epsilon_{ext}$ (which is the tolerated distance between the uniform distribution and the one of the final string) and on the particular algorithm adopted. 
In our case, we used the same implementation of \cite{code,framework}, which was proven to be a strong quantum proof extractor by \cite{De2012}. With respect to other implementations of the Trevisan extractor \cite{aaronson2018}, we require a longer seed, but we can extract a higher amount of random bits. Let us note that, since the length of the seed grows as $log(2n)^3$, where $n$ is the number of experimental runs, the randomness gain is not modified if we take also into account the bits invested in the classical extractor's seed. Indeed, the number of extracted bits grows polynomially in $n$. Hence, if $H_{min~Instr}^{\epsilon} \geq H_{min~CHSH}^{\epsilon}$, then $m_{Instr}-d_{Instr} \geq m_{CHSH}-d_{CHSH}$, where $m$ is the length of the final string (after the classical extraction) and $d$ the length of the required seed. For more details about the internal functioning of the classical randomness extractor and its specific parameter settings, see the Supplementary Information.

\section{Discussion}

In conclusion, in this work we implemented a device-independent random number generator based on the instrumental scenario. This shows that instrumental processes constitute an alternative venue with respect to Bell-like scenarios. Moreover, we also showed that, in some regimes, the efficiency of the randomness generated by the violation of the instrumental inequality \eqref{instrumental322} can surpass that of efficiency of the CHSH inequality, as shown in Fig.\ref{fig:gain}.

Through the proposed protocol, we could extract an overall number of 5270 random bits, considering a threshold for the instrumental violation of $I_{exp}=3.5$ and $10^{-1}$, both as error probability of the entropy accumulation protocol ($\epsilon_{EA}$), as well as smoothing parameter ($\epsilon$). The conversion rate, from public to private randomness, as well as the security parameters, could be improved on the same platform, by raising the number of invested initial random bits, or, analogously, the number of runs.
To access the regime in which the instrumental scenario is more convenient than the CHSH one, we should invest a number of random bits over $10^9$ and obtain a visibility of $\sim 0.98$ (the more we raise the amount of invested bits, the more the threshold for the visibility lowers down).

This proof-of-principle opens the path for further investigations of the instrumental scenario as a possible venue for other information processing tasks usually associated to Bell scenarios, such as self-testing \cite{Mayers04, Yang13, McKague14, Bamps15, Wu16, Supic16, Coladangelo17, McKague17, Supic18, Bowles18} and communication complexity problems \cite{Brukner04, Buhrman10, Buhrman16}.

\section{AKNOWLEDGEMENTS}
We aknowledge Gl\'aucia Murta for stimulating discussion. We acknowledge support from John Templeton Foundation via the grant Q-CAUSAL n$^{\circ}$61084 (the opinions expressed in this publication are those of the authors and do not necessarily reflect the views of the John Templeton Foundation). IA, DP, MM, GC and FS aknowledge project Lazio Innova SINFONIA. DC acknoledges a Ramon y Cajal fellowship, Spanish MINECO (Severo Ochoa SEV-2015-0522), Fundació Privada Cellex and Generalitat de Catalunya (CERCA Program). LG and LA acknowledge financial support from  the S\~ao Paulo Research Foundation (FAPESP) under grants 2016/01343-7 and 2018/04208-9. RC acknowledges the Brazilian ministries MCTIC, MEC and the CNPq (grants No. $307172/2017-1$ and $406574/2018-9$ and INCT-IQ) and the Serrapilheira Institute (grant number Serra-1708-15763). G.C. aknowledges Conicyt and Becas Chile. LA acknowledges financial support also from the Brazilian agencies CNPq (PQ grant No. 305420/2018-6 and INCT-IQ), FAPERJ (JCNE E-26/202.701/2018), CAPES (PROCAD2013 project), and the Brazilian Serrapilheira Institute (grant number Serra-1709-17173).

\section{METHODS}

\subsubsection{Experimental details}

Photon pairs were generated in a parametric down conversion source, composed by a nonlinear crystal beta barium borate (BBO) of 2 mm-thick injected by a pulsed pump field with $ \lambda = 392.5$ nm. 
After spectral filtering and walk-off compensation, photons of $ \lambda = 785$ nm are sent to the two measurement stations A and B.
The crystal used to implement active feed-forward is a $\mathrm{LiNbO}_{3}$ high-voltage micro Pockels Cell --made by Shangai Institute of Ceramics with $< 1$ ns risetime and a fast electronic circuit transforming each Si-avalanche photodetection signal into a calibrated fast pulse in the kV range needed to activate the Pockels Cell-- is fully described in \cite{sciarrino_teleportation}. 
To achieve the active feed-forward of information, the photon sent to Bob's station needs to be delayed, thus allowing the measurement on the first qubit to be performed. The amount of delay was evaluated considering the velocity of the signal transmission through a single mode fiber and the activation time of the Pockels cell. We have used a fiber 125 m long, coupled at the end into a single mode fiber that allows a delay of 600 ns of the second photon with respect to the first.

\end{document}


\title{Supplementary Information\\Experimental device-independent certified randomness generation with an instrumental causal structure}
\author{Iris Agresti}
\affiliation{Dipartimento di Fisica, Sapienza Universit\`{a} di Roma,
Piazzale Aldo Moro 5, I-00185 Roma, Italy}
\author{Davide Poderini}
\affiliation{Dipartimento di Fisica, Sapienza Universit\`{a} di Roma,
Piazzale Aldo Moro 5, I-00185 Roma, Italy}
\author{Leonardo Guerini}
\affiliation{International Center of Theoretical Physics - South American Institute for Fundamental Research, Instituto de F\'isica Te\'orica - UNESP, R. Dr. Bento T. Ferraz 271, 01140-070, S\~ao Paulo, Brazil}
\author{Michele Mancusi}
\affiliation{Dipartimento di Fisica, Sapienza Universit\`{a} di Roma,
Piazzale Aldo Moro 5, I-00185 Roma, Italy}
\author{Gonzalo Carvacho}
\affiliation{Dipartimento di Fisica, Sapienza Universit\`{a} di Roma,
Piazzale Aldo Moro 5, I-00185 Roma, Italy}
\author{Leandro Aolita }
\affiliation{International Center of Theoretical Physics - South American Institute for Fundamental Research, Instituto de F\'isica Te\'orica - UNESP, R. Dr. Bento T. Ferraz 271, 01140-070, S\~ao Paulo, Brazil}
\affiliation{Instituto de F\'{i}sica, Universidade Federal do Rio de Janeiro, Caixa Postal 68528, Rio de Janeiro, RJ 21941-972, Brazil}
\author{Daniel Cavalcanti}
\affiliation{ICFO - Institut de Ciencies Fotoniques, The Barcelona Institute of Science and Technology, E-08860 Castelldefels, Barcelona, Spain}
\author{Rafael Chaves}
\affiliation{International Institute of Physics, Federal University of Rio Grande do Norte, 59078-970, P. O. Box 1613, Natal, Brazil}
\affiliation{School of Science and Technology, Federal University of Rio Grande do Norte, 59078-970 Natal, Brazil}
\author{Fabio Sciarrino}
\affiliation{Dipartimento di Fisica, Sapienza Universit\`{a} di Roma,
Piazzale Aldo Moro 5, I-00185 Roma, Italy}

\maketitle

\section{1- NPA method applied to the  instrumental scenario}
In order to estimate the randomness generated in our experiment, we will adopt a similar strategy to that used in the Bell scenario (e.g. see \cite{rng_certified}). We assume that there exists an external observer, called Eve, that whats to predict the measurement outcomes $(a,b)$ of the instrumental test. The system used in the protocol is assumed to be quantumly correlated to a system held by Eve. In this scenario, the randomness in the outcomes $(a,b)$ is given by the min entropy of Eve's guessing probability (the probability that Eve correctly predicts the outomes).

In order to make this idea rigorous, let us first define the correlations that can be obtained the quantum instrumental scenario. Without loss of generality \cite{himbeeck2018, henson2014}, these can be expressed by the Born rule:
\begin{equation}\label{QInst}
    P_{Q}^{Inst}(a,b|x)=\tr(M_{a|x}\otimes N_{b|y=a} \rho_{AB}),
\end{equation} 
where $M_{a|x}$ and $N_{b|y}$ define proper quantum measurements (i.e. $M_{a|x}\geq 0$, $\sum_a M_{a|x}=\openone$,$N_{b|y}\geq 0$, $\sum_b N_{b|y}=\openone$) and $\rho_{AB}$ a bipartite quantum state (i.e. $\rho_{AB}\succeq0$ and $\tr \rho_{AB}=1$).
 In the randomness analysis, we assume that additionally to systems $AB$ there exists a third party, Eve, that is trying to guess the outcomes $a$ and $b$. The full picture is described by a tripartite  state $\ket{\Psi_{ABE}}$, onto which local measurements are performed (notice that the state can be assumed to be pure since the dimension of the systems are considered arbitrary).
 The statistics obtained is then evaluated on an instrumental functional $\mathcal{I}$ (e.g. the left-hand-side of inequality 1 of the main text),  which allows us to calculate Eve's  maximum guessing probability. Eve is assumed to know $\ket{\Psi_{ABE}}$ and the measurement implemented at $AB$. 
 
 In this case, the maximum probability  that Eve guesses correctly $(a, b)$ for a fixed setting  $x = j$, and given that $\mathcal{I}=\beta$, can be written as follows:
 \begin{align*}
 \max&\quad \sum_{a,b}\langle\Psi_{ABE}|\Pi_{a|x=j}\otimes\Pi_{b|y=a}\otimes\Pi_{e= (a,b)}|\Psi_{ABE}\rangle \\
 \textrm{s.t.}&\quad P(a,b|x) =  \langle\Psi_{ABE}|\Pi_{a|x}\otimes\Pi_{b| y=a}\otimes\mathbb{I}_E|\Psi_{ABE}\rangle ,\\& \quad \forall a,b,x \quad  \mathcal{I}(\{P(a,b|x)\})=\beta,
 \end{align*}
 where the maximization is taken over all  possible tripartite states $\Psi_{ABE}$  and local measurements  $\{\Pi_{a|x}\},\{\Pi_{b|y}\},\{\Pi_e\}$. In other words, we want the maximum probability that Eve's outcome $e$ equals $(a,b)$, among all tripartite quantum correlations satisfying $\mathcal{I}=\beta$. The amount of randomness (in bits) is given by $H_\infty(A,B|E,x=j)=-\log_2(P_{guess}(x= j))$, where $P_{guess}(x= j)$ denotes the solution of the previous optimization problem.
 
 This optimization problem is computationally intractable, as it considers quantum systems of arbitrary dimension. A way out is to upper-bound its value by  using the Navascu\'es-
 Pironio-Ac\'in (NPA) hierarchy  \cite{bound}. The NPA  hierarchy is a numerical tool used to generate a sequence of sets of correlations $\mathcal{Q}_1 \supset  \mathcal{Q}_2 \supset \ldots \mathcal{Q}_n $  that converges to the set of quantum correlations of any bipartite system in the Bell scenario, defined as those that can be written as 
 \begin{equation}\label{QBell}
     P^{Bell}_Q(a,b|x,y)=\tr(M_{a|x}\otimes N_{b|y} \rho_{AB}),
 \end{equation} 
 where, as previously mentioned, $M_{a|x}$ and $N_{b|y}$ define proper quantum measurements $\rho_{AB}$ a bipartite quantum state. Notice that, by definition, $P_{Q}^{Inst}(a,b|x)=P^{Bell}_Q(a,b|x,y=a)$. We can now use the NPA hierarchy to define the following optimization problem that upper bounds the guessing probability of Eve in an instrumental scenario:
 \begin{align}
 \label{eq:randomnessSDP}
 P_{guess}^k(x=j) =&\max_{P(a,b,e|x,y)}\quad \sum_{a,b}  P(a,b,e=(a,b)|x=j,y=a)\\ \nonumber
 &\textrm{s.t.}\quad  \{P(a,b|x,y,e=e_0)\}\in\mathcal{Q}_k,\  \forall e_0\\ \nonumber
 &\quad \mathcal{I}(\{\sum_eP(a,b,e|x,y=a) \})=\beta.
 \end{align}
 This problem maximises the probability that Eve's outcome $e$ equals $(a,b)$, among all of the possible correlations achieving the observed instrumental violation value $\beta$ and for which the conditional correlations within the systems $AB$, conditioned to Eve's outcomes, belong to $Q_k$.
 Notice that, since for every $k$ the set of correlations over which we perform our optimization is bigger than the quantum one, $P_{guess}^k(x=j)\geq P_{guess}(x=j)$, the solution of \eqref{eq:randomnessSDP} provides a lower bound to the randomness obtained. In our implementation, we used $I$ as the left-hand-side of inequality 1 of the main text, and $k=2$.
 
 The optimal value of the  above problem is used to define  $f_{x=j}(\mathcal{I})=-\log_2(P_{guess}^k(x= j))$ in the lower bound  $H_\infty(A,B|E,x=j)\geq  f_{x=j}(\mathcal{I})$ mentioned in the  main text.

\section{2- Entropy Accumulation Theorem in the Instrumental scenario}
 
In order to make our protocol secure 
 against general quantum adversaries, we adapt the techniques developed by \cite{friedman2019} to our scenario.
 This method, as mentioned in the main text, resorts to the \textit{Entropy Accumulation Theorem} (EAT), in order to deal with processes not necessarily made of independent and identically distributed (\textit{iid}) runs. 
 This tool ensures that our protocol is secure, even if an adversary has access to the devices and introduces a (quantum) memory inside, making each run possibly depend on the previous ones. In this paragraph, we briefly recall the EAT and then explain in further details how we applied it to our case.
 
  The EAT provides a lower bound on the smooth min-entropy of any process that can be described as a sequential application of $n$ \textit{EAT channels}. This bound is polynomial in the number of runs $n$ and, when the number of runs grows, it tends to that obtaining assuming \textit{iid}.
 The $i$-th EAT channel ($\mathcal{M}_{i}$) can be described as follows: it outputs two registers $O_{i}$, which are given to the honest parties Alice and Bob and that represent the information that should be kept secret, another system $S_{i}$ constituting some side-information leaked by the channel, and a memory $R_{i}$, which is passed as input to the next map ($\mathcal{M}_{i+1}$). 
 Also a system E is considered in the model, representing any additional side information correlated to the initial state. Hence the systems $O$ are the ones in which the entropy is accumulated, conditioned on S and E. 
We also consider, for each round, a classical value $C_{i}$ which is evaluated from $O_{i}$ and $S_{i}$. Hence, $\mathcal{M}_{i}: R_{i-1} \to R_{i}O_{i}S_{i}C_{i}$ for $i \in [n]$.
A subset $c_{1}^{n} \in C^{n}$, where $C^{n}$ is the space of all possible n-bit sequences, defines an event $\Omega$ and we indicate with $\hat{\Omega}$ the empirical statistics of $c_{1}^n$, i.e. $\hat{\Omega}=\{freq(c_{1}^n)| c_{1}^n \in \Omega\}$.

 The conditions that a EAT channel must fulfil are the following (for a formal definition, see the Methods section of \cite{friedman2019}):
 \begin{itemize}
     \item $C_{i}$ are classical systems of finite-dimension (random variables), while $O_{i}$, $R_{i}$, and $S_{i}$ are quantum registers. $d_{O_{i}}$ is the maximum dimension of $O_{i}$;
     \item for any input state $\sigma_{R_{i-1}R'}$, where $R'$ is isomorphic to $R_{i-1}$, the output state $\sigma_{R_{i}O_{i}S_{i}C_{i}R'}=(\mathcal{M}_{i} \otimes I_{R'})\sigma_{R_{i-1}R'}$ has the property that $C_{i}$ can be evaluated from $\sigma_{O_{i}S_{i}}$ without changing the state;
     \item (Markov chain condition) for any initial state $\rho^0_{R_{0}E}$, the final state $\rho_{O_{1}^n S_{1}^n C_{i}^n E}= Tr_{R^{n}}(\mathcal{M}_{n}\otimes \mathcal{M}_{n-1} ... \mathcal{M}_{1} \otimes I_{E})\rho_{R_{0}E}^0$ fulfils the Markov chain condition $O_{1}^{i-1} \leftrightarrow{} S_{1}^{i-1}E \leftrightarrow{} S_{i}$ for each $i \in [n]$. In other words, $I(O_{1}^{i-1}:S^{i}|S^{i-1} E)=0$, so the previous outcomes $O_{1}^{i-1}$ are independent of future side information $S_{1}^{i}$, given all the past side information $S_{1}^{i-1}E$.
 \end{itemize}
 
 Now, we state the EAT, as in \cite{friedman2019}:\\
\textbf{Entropy Accumulation Theorem (EAT).} Let $\mathcal{M}_{i}: R_{i-1} \to R_{i}O_{i}S_{i}C_{i}$ be the for $i \in [n]$ be EAT channels, $\rho$ be the final state, $\Omega$ an event defined over $C^n$, $p_{\Omega}$ the probability of $\Omega$ in $\rho$, and $\rho_{|\Omega}$ the final state conditioned on $\Omega$ and let $\epsilon \in (0, 1)$. For $f_{min}$ a tradeoff function for $\{\mathcal{M}_{i}\}$, $\hat{\Omega}=\{freq(c_{1}^n)|c_{1}^n \in \Omega\}$ convex and any $t \in \mathcal{R}$, such that $f_{min}(freq(c_{1}^n))\geq t$ for any $c_{1}^n \in C_{1}^n$ for which $P(c_{1}^n)_{\rho_{|\Omega}}\geq0$,
\begin{equation}
    \mathcal{H}_{min}^{\epsilon}(O^{n}|S^n E^n) > nt-\nu
    \sqrt{n}
    \label{newbound}
\end{equation}
where $t(\mathcal{I})|_{\mathcal{I}_{exp}}$ represents the Von-Neumann entropy, which depends on the extent of the violation that would be obtained from an honest implementation of the protocol, i.e. with no adversary, $I_{exp}$. 
On the other hand, $\nu$ depends on $||\nabla{f_{min}}||_{\infty}$, on the smoothing parameter $\epsilon$ and on $\epsilon_{EA}$, which can be interpreted as the error probability of the entropy accumulation protocol.

\begin{figure}[H]
\centering
\includegraphics[scale=0.69]{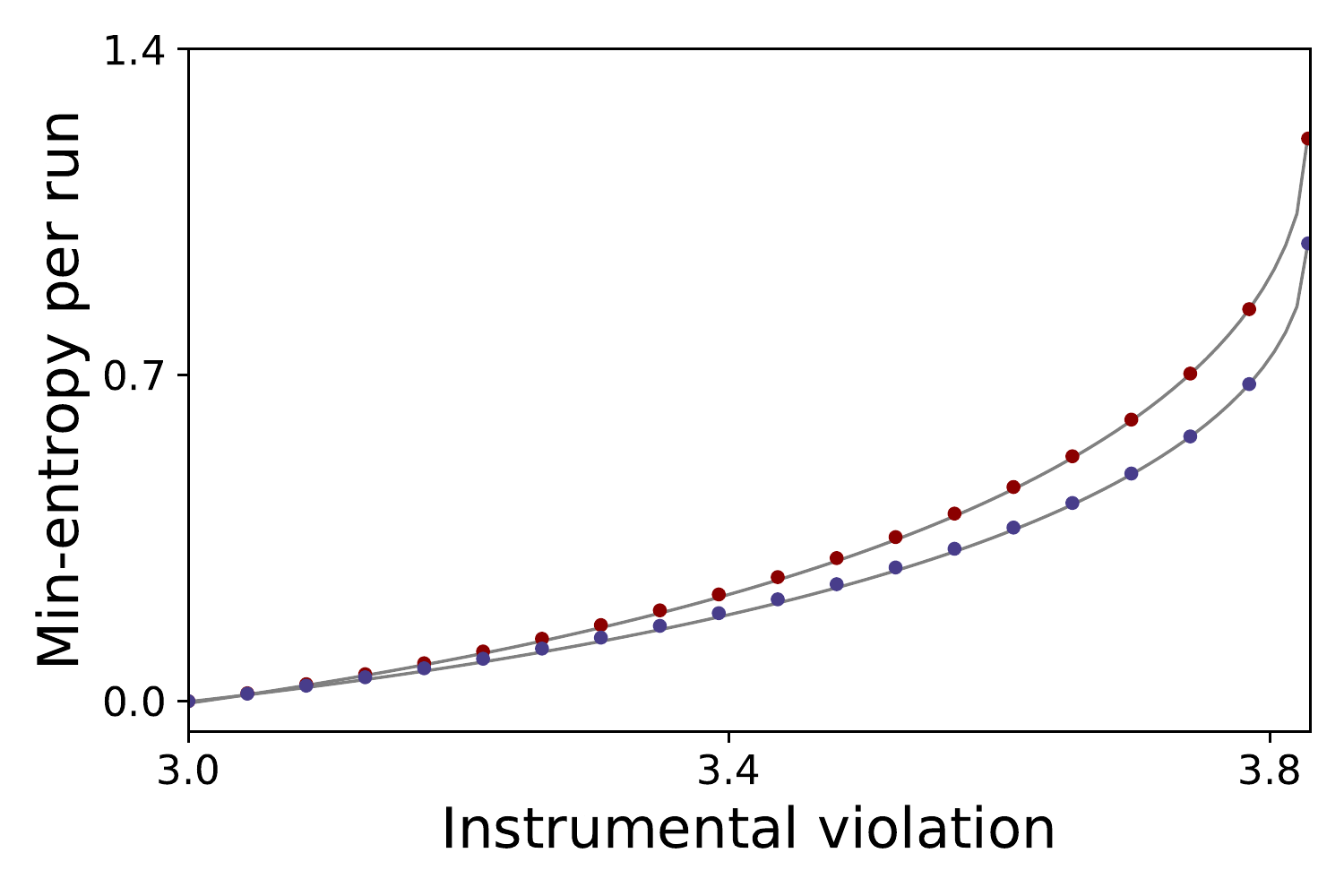}
\caption{\textbf{Lower bounds of $f_{x}(\mathcal{I})$.} In this plot, we present the $f_{x}(\mathcal{I})$ min-entropy lower bounds 
obtained fitting the numerical data given by the techniques of \cite{rng_certified, bound}. The red dots correspond to $x=2,3$, while the blue ones to $x=1$. The grey curves represent the lowered best fit functions $f_{1}'= u_{1}'-w_{1}\mathcal{I}-z_{1}\sqrt{1+2\sqrt{2}-\mathcal{I}}$ (lowest curve) and $f_{2}' = f_{3}' = u_{2,3}'-w_{2,3}\mathcal{I}-z_{2,3}\sqrt{1+2\sqrt{2}-\mathcal{I}}$ (highest curve). We have $(u_{1}', w_{1}, z_{1})=3.062, 0.542, 1.579)$ and $(u_{2,3}', w_{2,3}, z_{2,3})=(3.735, 0.657, 1.942)$.
}
\label{fig:fits}
\end{figure}

 In order to adapt the protocol of \cite{friedman2019} to our case, we first of all need to build a $f_{min}(p(a,b|x))$ function for the instrumental process and evaluate the smooth min-entropy bound according to inequality (S2). In order to perform the required derivatives, we fit the data computed via SDP (see previous section), as shown in Fig.\ref{fig:fits}. The obtained fits $f_{x}(\mathcal{I})$, for $x=1,2,3$, are the following:
\begin{equation}
    f_{1} = u_{1}-w_{1}\mathcal{I}-z_{1}\sqrt{1+2\sqrt{2}-\mathcal{I}}
    \label{eq_f1}
\end{equation}
and 
\begin{equation}
f_{2} = f_{3} = u_{2,3}-w_{2,3}\mathcal{I}-z_{2,3}\sqrt{1+2\sqrt{2}-\mathcal{I}}
\label{eq_f23}
\end{equation}

where $(u_{1}, w_{1}, z_{1})=(3.067, 0.542, 1.579)$ and $(u_{2,3}, w_{2,3}, z_{2,3})=(3.740, 0.657, 1.942)$. To ensure that we will not overestimate the min-entropy, we lower these curves along the y-axis, in order to give a lower bound on the numerically evaluated min-entropies (see Fig. \ref{fig:fits}). Hence, the parameters $u_{1}'$ and $u_{23}'$ that we will use in our calculations will be respectively $u_{1}'= 3.062$ and $u_{23}'= 3.735$.

Then, we can implement the protocol, as follows: we group our experimental runs in $k$ blocks, with no fixed length; at each run, a binary variable $T$ is evaluated, in particular, it is 1 with an arbitrary probability $\gamma$. Each block ends either when $T=1$, or when $T=0$ for $s_{max}$  times in a row (this parameter is set by the user). So, after $n$ runs, we will have $k$ blocks, each one containing at most 1 test run. Each of these blocks corresponds to a EAT channel $\mathcal{M}_{i}$ and we consider that $O_{i} \rightarrow \vec{A}_{i}\vec{B}_{i}$ and $S_{i} \rightarrow \vec{X}_{i}\vec{T}_{i}$, where $\vec{A}_{i}$ and $\vec{B}_{i}$ are classical registers containing the string that Alice and Bob get throughout the $i-th$ block; on the other hand, $\vec{X}_{i}$ and $\vec{T}_{i}$ are the strings of Alice's inputs and of the $T$ values for the same block. In the end, $C_{i}$ is the tuple ($a_{i}$, $b_{i}$, $x_{i}$), obtained still in the $i-th$ block, for $T=1$, i.e. in the test run. 

The length of the blocks (as well as that of $\vec{A}_{i}$, $\vec{B}_{i}$, $\vec{X}_{i}$ and $\vec{T}_{i}$) is not fixed, but the expected value is the following: $s'=\frac{1-(1-\gamma)^{s_{max}}}{\gamma}$.
Now, let us check whether the aforementioned channel $\mathcal{M}_{i}: R_{i-1} \to R_{i}\vec{A}_{i}\vec{B}_{i}\vec{X}_{i}\vec{T}_{i}C_{i}$, is indeed an EAT channel.
\begin{itemize}
    \item $C_{i} \in \{\{0,1,-\}, \{0,1,-\}, \{0,1,2,-\}\}$ for any $i \in [1,k]$, where $\{-,-,-\}$ is the case in which the $i-th$ block ends with no test run, and $d_{O_{i}} \leq 6^{s_{max}}$.
    \item $C_{i}$ is evaluated by the classical registers $\vec{A}_{i}$, $\vec{B}_{i}$, $\vec{X}_{i}$ and $\vec{T}_{i}$, so it does not modify the marginals on those registers.
    \item the Markov chain condition holds, provided that the inputs are chosen independently at each round. In this case, the condition $\vec{A}_{1, ..., i-1}\vec{B}_{1, ..., i-1} \leftrightarrow \vec{X}_{1, ..., i-1}\vec{T}_{1, ..., i-1}E \leftrightarrow \vec{X}_{i}\vec{T}_{i}$ trivially holds.
\end{itemize}
and the min-entropy in the $i$-th block will be bounded as follows (for the demonstration, see Methods section of \cite{friedman2019}): 
	\begin{equation}
	H(\vec{A}_{i}, \vec{B}_{i}| \vec{X}_{i}, \vec{T}_{i}, E) > (s'-1). f_{2,3}'(\mathcal{I}) +f_{1}'(\mathcal{I})
	\label{eq:me}
	\end{equation}
In the inequality (\ref{eq:me}), we are considering that, for $x=2,3$ and $x=1$, the bounds are different. Hence, in the accumulation rounds, we consider $f_{2,3}$, while, for the test round, we take $f_{1}$, i.e. the lowest bound.
The relation (\ref{eq:me}) holds when the min-entropy is evaluated on a state conditioned on the event $\Omega$ that, in our case, consists in obtaining an instrumental violation value higher than the treshold $I_{exp}$, throughout the test runs. In other words, given the input $x_{i}$ of the $i-th$ test run, the outputs ($a_{i}$ and $b_{i}$) follow the joint probability distribution $P(a,b|x, T=1)=\frac{3 p(a, b, x, T=1)}{1-(1-\gamma)^{s_{max}}}$, with corresponding violation $\mathcal{I}(\frac{3 p(a, b, x, T=1)}{1-(1-\gamma)^{s_{max}}})= \frac{3 I^{*}(p(a, b, x, T=1))}{1-(1-\gamma)^{s_{max}}} = I_{exp}$. Let us note that $\sum_{a,b,x}p(a, b, x, T=1)=1-(1-\gamma)^{s_{max}}$.

\begin{figure}[t]
\centering
\includegraphics[scale=0.69]{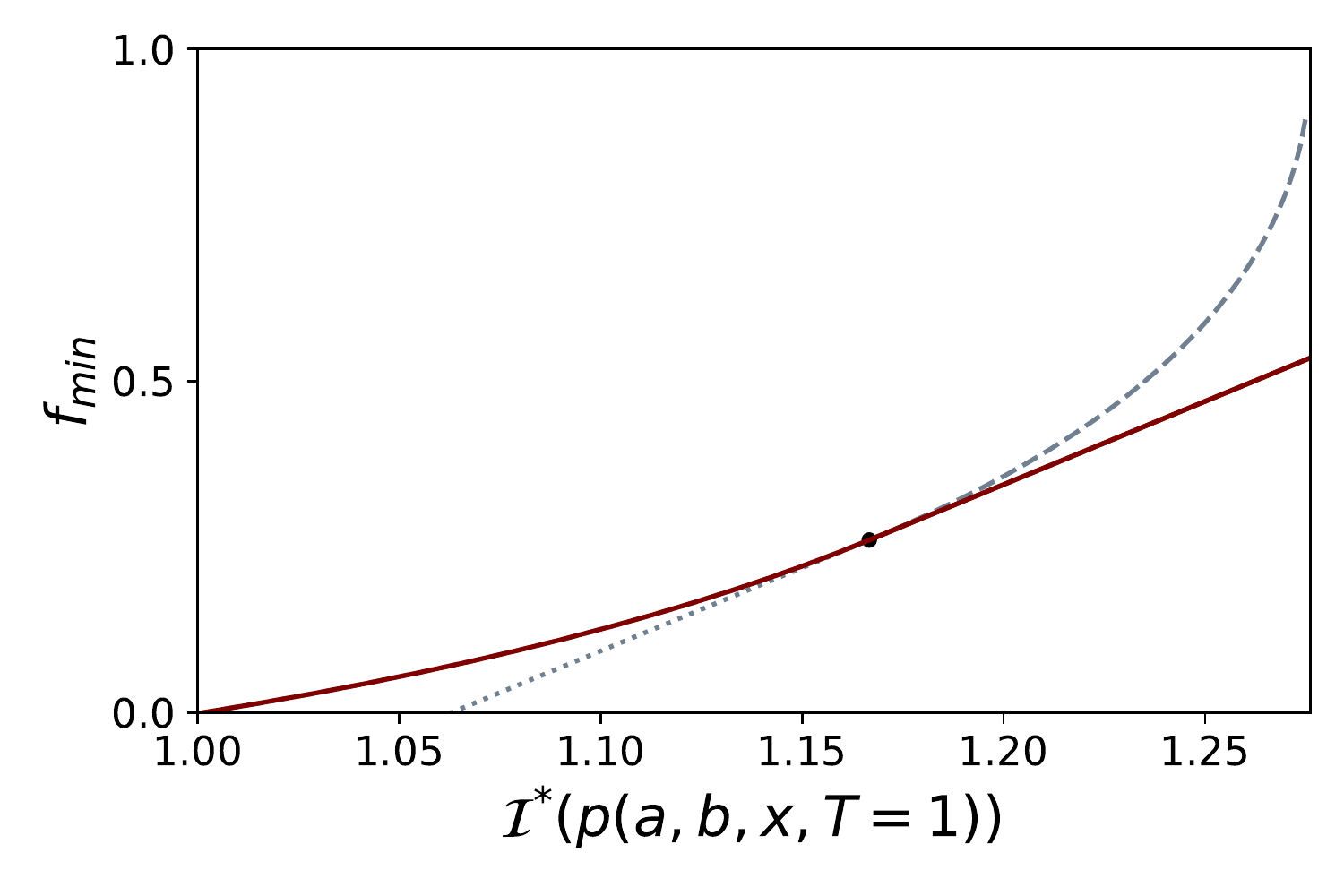}
\caption{ \textbf{Tradeoff function $f_{min}$.}  In this plot, we show the construction of the min-tradeoff function $f_{min}(I^{*}_{t}(p(a,b,x,T=1)))$. To ensure that its derivative does not explode, we cut the function in an arbitrary point and glue it to a linear function.}
\label{fig:lin}
\end{figure}

At this point, we define
 \begin{equation}
g(\mathcal{I}^{*})=(s'-1). f_{2,3}(\mathcal{I}) +f_{1}(\mathcal{I})
\label{eq_g}
\end{equation} 
and, substituting (\ref{eq_f1}) and (\ref{eq_f23}) in (\ref{eq_g}), we obtain:
\begin{equation}
g(\mathcal{I}^{*})=(u_{1}'+(s'-1)u_{2,3}')-3(w_{1}+(s'-1)w_{2,3})\frac{\mathcal{I}^{*}}{1-(1-\gamma)^{s_{max}}}-(z_{1}+(s'-1)z_{2,3})\sqrt{1+2\sqrt{2}-3\frac{\mathcal{I}^{*}}{1-(1-\gamma)^{s_{max}}}}
\end{equation} 
The $f_{min}$ convex tradeoff function will be defined as:
 \begin{numcases}{f_{min}(\mathcal{I}^{*}(a, b, x, T=1), \mathcal{I}^{*}_t)=}
 g(\mathcal{I}^{*}), & for $\frac{3 I^{*}(p(a, b, x, T=1))}{1-(1-\gamma)^{s_{max}}} \in (0, \mathcal{I}_t)$\\
 \frac{d g(\mathcal{I}^{*})}{d\mathcal{I}^{*}}|_{\mathcal{I}^{*}_t} (\mathcal{I}^{*}-\mathcal{I}^{*}_t)+ g(\mathcal{I}^{*}_t), & for $\frac{3 \mathcal{I}^{*}(p(a, b, x, T=1))}{1-(1-\gamma)^{s_{max}}} \in (\mathcal{I}_t, 1+2\sqrt{2})$
  \end{numcases}
  Concatenating the function $g(\mathcal{I}^*)$ to a linear function in a given point $\mathcal{I}_t=\frac{3 \mathcal{I}^{*}_t}{1-(1-\gamma)^{s_{max}}}$, as shown in Fig. \ref{fig:lin}, ensures that the derivative of $f_{min}$, which contributes to the right term of inequality (\ref{newbound}), does not explode.
 
In the end, for a given $\mathcal{I}_{exp}$, as proven in \cite{friedman2019}, the protocol either aborts with probability higher than $1-\epsilon_{EA}$, or it gives at least the following smooth min-entropy:
	\begin{equation}
	    \frac{n}{s'}\eta_{opt}(\epsilon, \epsilon_{EA})
	\end{equation}
where $\eta(I^{*},I^{*}_t,\epsilon, \epsilon_{EA})= f_{min}(I^{*},I^{*}_t)-\sqrt{s/n} 2 (log(1+2 \times 6^{smax})+ 4 ||\frac{d g(I^{*})}{dI^{*}} ||_{I^{*}_t})\sqrt{1-2 log(\epsilon\epsilon_{EA})}$ and $\eta_{opt}(\epsilon, \epsilon_{EA})= max_{(I^{*}_t)} \eta(I (1-(1-\gamma^{smax}))/3-4\delta', I^{*}_t,\epsilon, \epsilon_{EA})))$.
	
\begin{figure*}[t]
\includegraphics[scale=0.69]{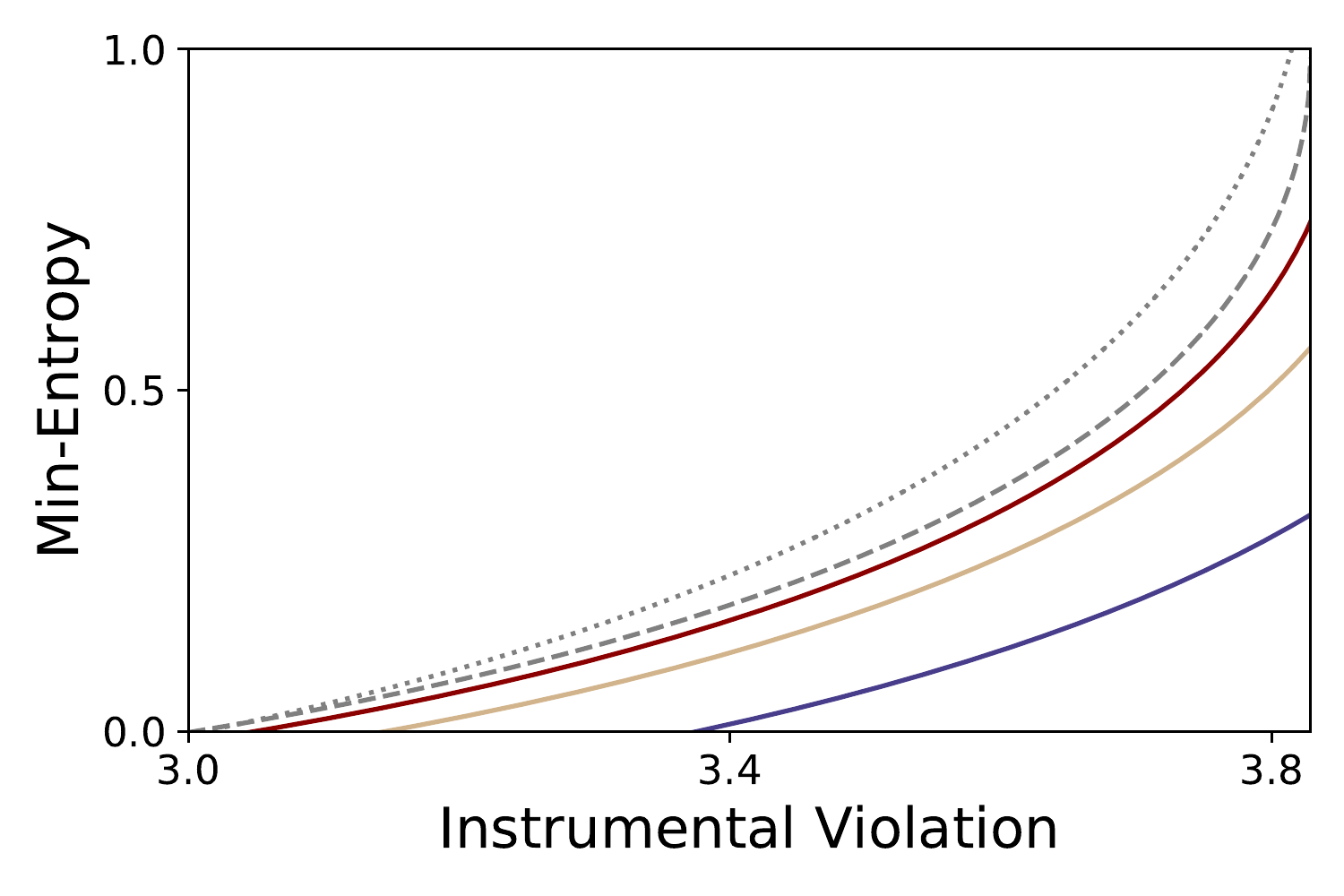}
\caption{\textbf{Smooth min-entropy lower bound given by EAT.} In this plot, we show the lower bound, guaranteed by the EAT, on the smooth min-entropy, in terms of the expected instrumental violation $I_{exp}$ for the different ($\epsilon$, $\epsilon_{EA}$, $\delta'$, $n$) parameters' choices. The red curve is obtained with ($10^{-6}$, $10^{-6}$, $10^{-3}$, $10^{8}$), the golden curve is obtained with ($10^{-6}$, $10^{-6}$,$10^{-3}$, $10^{7}$) and  the blue curve is obtained with($10^{-5}$, $10^{-5}$, $ 10^{-3}$, $10^{6}$). The dotted curve represents the \textit{iid} bound for $x=2,3$, while the dashed one is for $x=1$.}
\label{etaopt}
\end{figure*}

The $\eta_{opt}$ bound is plotted in terms of the expected violation of the instrumental inequality, for different parameter settings in Fig. \ref{etaopt}.

The protocol described above is a $(\epsilon_{c}, \epsilon_{s})-$secure protocol, where $\epsilon_{c}$ and $\epsilon_{s}$ indicate, respectively, the \textit{completeness} and the \textit{soundness} of the protocol, which are defined as follows \cite{shen2018, friedman2019,Friedman2017}:
\begin{itemize}
    \item \textbf{Soundness}: in its implementation the protocol either aborts or returns an $m-$bit string $Z \in \{0,1\}^m$, which is characterized by $(1-P(abort))||\rho_{ZRE}-\rho_{U_{m}}\otimes\rho_{U_{R}}\otimes\rho_{E}||\leq \epsilon_{s}$
    where $R$ is the input randomness register, $E$ is the adversary system, and $\rho_{U_{R}}$ and $\rho_{U_{m}}$ are completely mixed states on the appropriate registers.
    \item \textbf{Completeness}: there exists an honest implementation which aborts with probability less than $\epsilon_{c}$
\end{itemize}
In our case, $\epsilon_{c}$ and $\epsilon_{s}$ are quantified as follows. $\epsilon_{s}$ is made of several contributions \cite{shen2018, friedman2019,Friedman2017}, in particular we have to take into account 3 terms: (i) the $\epsilon_{ext}$, which is the internal error of the classical extractor, described in Par. \textit{Randomness Extractor} of this supplemental material, (ii) the parameter $\epsilon$ characterizing the smooth min-entropy $\mathcal{H}^{\epsilon}_{min}$ and (iii) the probability of the protocol not aborting ($\epsilon_{EA}$). Hence $\epsilon_{s}=\epsilon_{EA}+m\epsilon_{ext}+\frac{\epsilon}{2}$. In our experimental condition, the resulting soundness amounts to $\epsilon_{s} \sim 10^{-1}$. 
On the other hand, the completeness can be evaluated by the Hoeffding's inequality \cite{hoeffding1963} (since it refers a \textit{iid} honest implementation), as $P(abort) \leq e^{-2n\delta'^2}= \epsilon_{c}$, so it depends on the experimental uncertainty on the violation. In our case, $\epsilon_{c} \sim 10^{-37}$.\\
Note that, if $\gamma \neq 1$, both the soundness and completeness get an extra-contribution, given by  the maximum statistical distance between the distribution of the drawn $\{T_{1}, ..., T_{n}\}$ from the ideal one, i.e. that of $n$ \textit{iid} random variables drawn according to the Bernoulli distribution, with parameter $\gamma$, if the \textit{interval algorithm} \cite{interval_algorithm} is adopted. 
 
\section{3- Randomness Extractor}
\label{extractor}
Quantum correlations can be exploited to generate random numbers, whose randomness can be device-independently certified. However, in real experimental conditions, quantum correlations are inevitably mixed with classical noise, which could be controlled and used by an adversary (Eve) to gain partial information about the output random bits. This is the reason why there is the need to apply a post-processing procedure to filter the true randomness out of the raw bits generated by a Quantum Random Numbers Generator (QRNG). This procedure is called \textit{randomness extraction} and it involves the use of classical algorithms known as \textit{randomness extractors} \cite{new, introduction}. The aim of these algorithms is to obtain a bit sequence, following an almost uniform distribution, up to an error parameter $\epsilon_{ext}$, which can be made arbitrarily small.
The inputs of a randomness extractors are the following: a weak random source, constituted by the raw bits and characterized by a min-entropy of at least $k$, and a seed of length $d$, not necessarily uniform \cite{non uniforme}. The accuracy of the extractor can be increased, for a given min-entropy $k$, by reducing the number of extracted bits and by injecting a longer seed.

A deterministic function which takes as inputs the source and the seed and achieves these goals is called $(k,\epsilon_{ext})-extractor$. The optimum situation, therefore, would be to have $m$, the number of extracted bits, as close to $k$ as possible, meaning that all the available min-entropy has been extracted (indeed $k-m$ is the \textit{entropy loss}), and the seed, of length $d$, as small as possible, to minimise the amount of additional randomness.

 Recently, an important and promising randomness extractor, Trevisan's extractor \cite{trevisan}, has aroused considerable theoretical interest, since it has been proven to be secure against quantum adversaries \cite{non uniforme}. This algorithm requires a seed that scales polylogarithmically in the input size, giving an advantage over extractors adopting (almost) universal hashing, whose seed grows polinomially with the length of the input. Furthermore, Trevisan's extractor is also proven to be a strong extractor \cite{strong proof}, i.e. the seed can be concatenated to the extracted string and used as a part of the result, being nearly independent from the output string. Two implementations of this extractor were made by Ma \textit{et al.} \cite{post} and Mauerer \textit{et al.} \cite{framework}.
 
Trevisan's construction, in a few words, is a recipe to build a randomness extractor combining two elements: (i) the \textit{weak design}, that divides the initial seed into smaller blocks of random bits of length $t$ (\textit{sub-seeds}) and (ii) the \textit{one bit extractor}, which extracts a single random bit from the random source for each block.
In the \textit{weak design}, the blocks $\{S_{1},..., S_{j}\}$ into which the seed is divided should be nearly independent to ensure that the maximum amount of entropy is extracted. Hence, a family of sets $S_{1},\ldots S_{m} \subset [d]$ is a \textit{weak (m, t, r, d)-design} if\\
1. For all $i$, $|S_{i}|=t$\\
2. For all $i$, $\sum_{j=1}^{i-1} 2^{|S_{j}\cap S_{j }|} \leq rm$, where the parameter $r$ is the so-called \textit{overlap} of the weak design.\\
Each of the $S_{j}$ is fed into a one bit extractor and they are all in the end concatenated into a final random string.

In our work, we adopt the so-called \textit{Block weak design}, which is a variation of another algorithm, which we will refer to as the \textit{Standard weak design}. The latter is a refined version of Nisan and Wigderson \cite{Nisan}, whose effectiveness was proved by Hartman and Raz \cite{Hartman}, under the parameters choice given by $r=2e$ and $d=t^2$ with $t = 2\lceil\log n + 2 \log 2/\epsilon_{ext}\rceil$, where $n$ is the number of runs. The \textit{Block weak design}, on the other hand, was developed by Ma and Ta \cite{Ma} modified by Mauer \textit{et al.} \cite{framework} with $r=1$ and $d=(l+1)t^2$, where $l:= \text{max}\{1,\lceil\frac{\log(m-r')-\log(t-r')}{\log(r')-\log(r'-1)}\rceil\}$ and $r'=2e$. In comparison, the \textit{Block weak design} requires a seed's length exceeding that of the input weak random source's string, but it allows to extract more bits from the source, due to a smaller $r$, with respect to the \textit{Standard weak design}.
The one-bit extractor is realized by an error correcting code, which is constructed by concatenating a Reed-Solomon code with an Hadamard code.
Hence, as a preliminary step, we fix the following three parameters: (i) $2n$ (input length), (ii) $\alpha$ (min-entropy per bit, certified by the experimental instrumental violation) and (iii) $\epsilon_{ext}$ (error per bit). 
After that, we derive the seed length, the total min-entropy $k=\mathcal{H}_{min}^{\epsilon}$ and $m=(k-4\log\frac{1}{\epsilon_{ext}}-6)/r$. As we can see, fixing the lenght $m$ of the output string and the error per bit $\epsilon_{ext}$,
the required min-entropy increases with the overlap of the sets $\{S_i\}$.

\begin{figure}[H]
\centering
\includegraphics[scale=0.69]{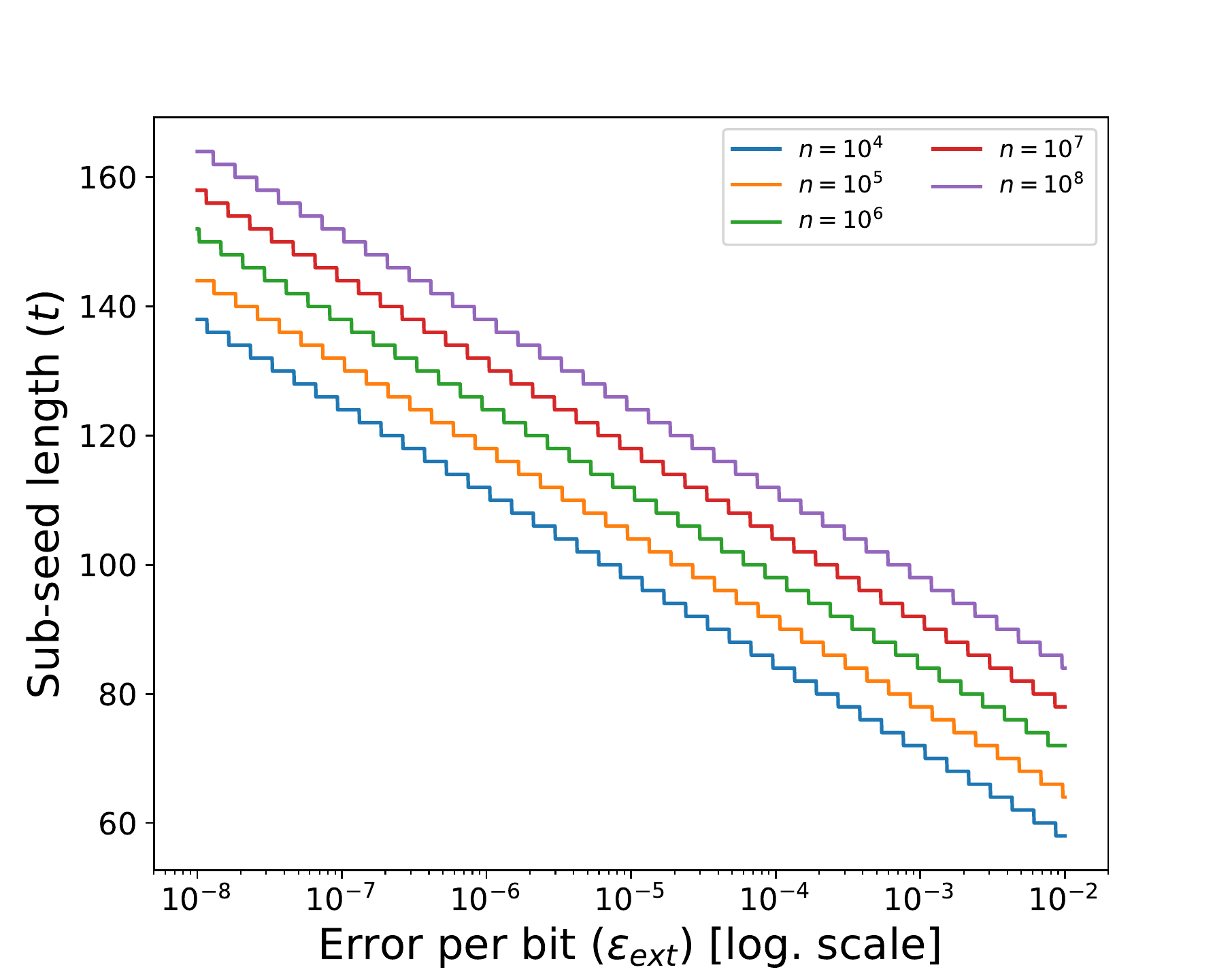}
\caption{\textbf{Length of the sub-seeds produced by the weak design depending on the error per bit.} This plot shows how the sub-seed's length produced by the weak design, i.e. the small sets into which the weak design algorithm divides the seed, changes in terms of the error per bit $\epsilon_{ext}$. It can be noted that in semi-log scale, the sub-seed's length $t$ is a step function and decreases linearly as the error per bit $\epsilon_{ext}$ increases. Furthermore, the greater the input $n$, the greater the length of the single sub-seed created by the weak design. This plot refers both to the Standard weak design \cite{Nisan} as well as the Block weak \cite{framework} design (which we adopt in our paper) algorithms.}
\label{fig:extr_1}
\end{figure}

\begin{figure}[H]
\centering
\includegraphics[scale=0.69]{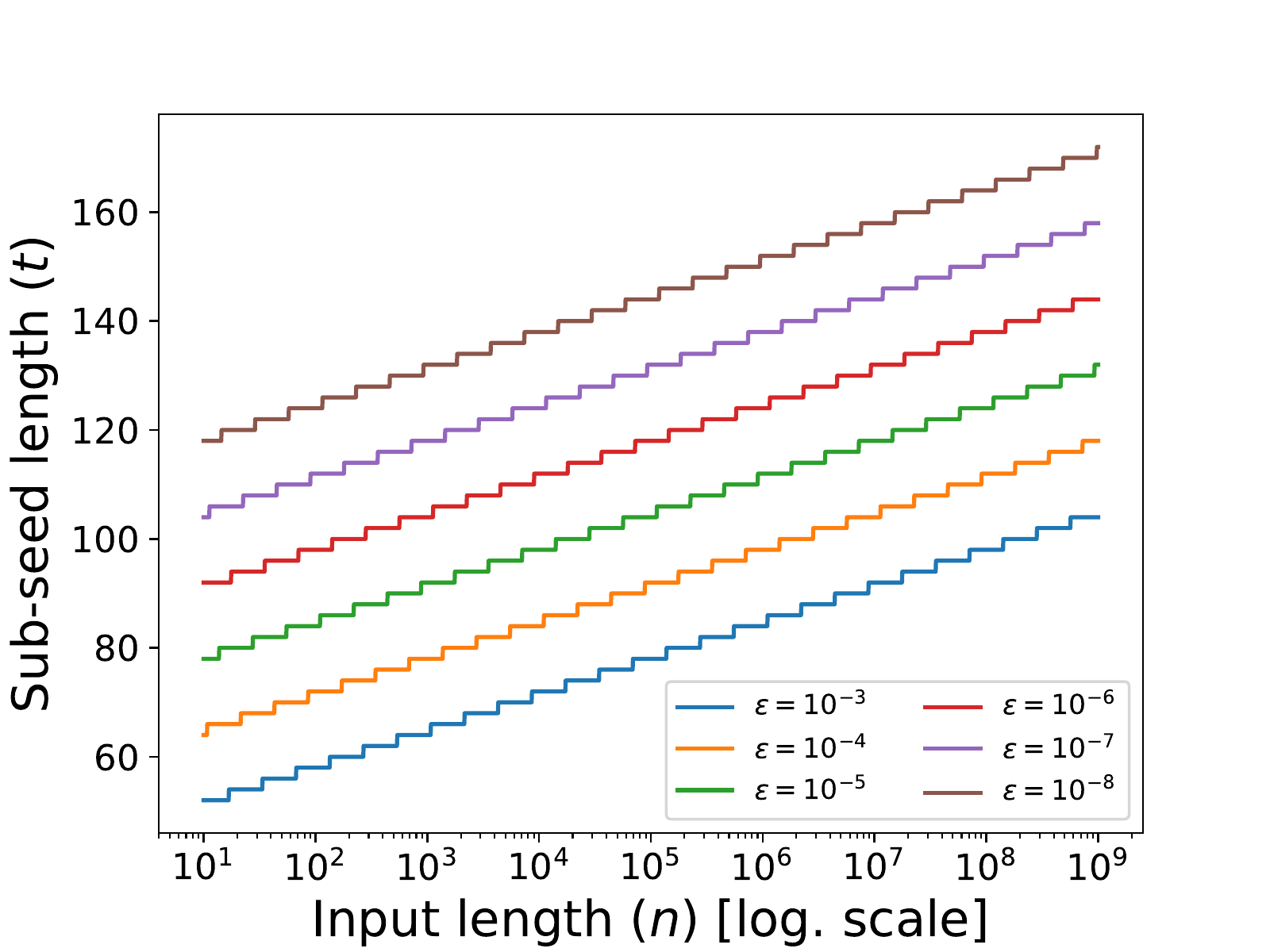}
\caption{\textbf{Length of the  sub-seeds produced by the weak design depending on the input length of the source.} In this figure is represented how the sub-seed length $t$, i.e. the small sets into which the weak design algorithm divides the seed, varies as a function of the input length $n$, plotted for different error per bit $\epsilon_{ext}$ parameters. We can see that in semi-log scale the sub-seed length is a step function and increases linearly with the input. Furthermore, the greater the error per bit, the smaller will be the length of the single sub-seed created by the weak design. This plot refers both to the Standard weak design \cite{Nisan} as well as the Block weak \cite{framework} design (which we adopt in our paper) algorithms.}
\label{fig:extr_2}
\end{figure}

\begin{figure}[H]
\centering
\includegraphics[scale=0.69]{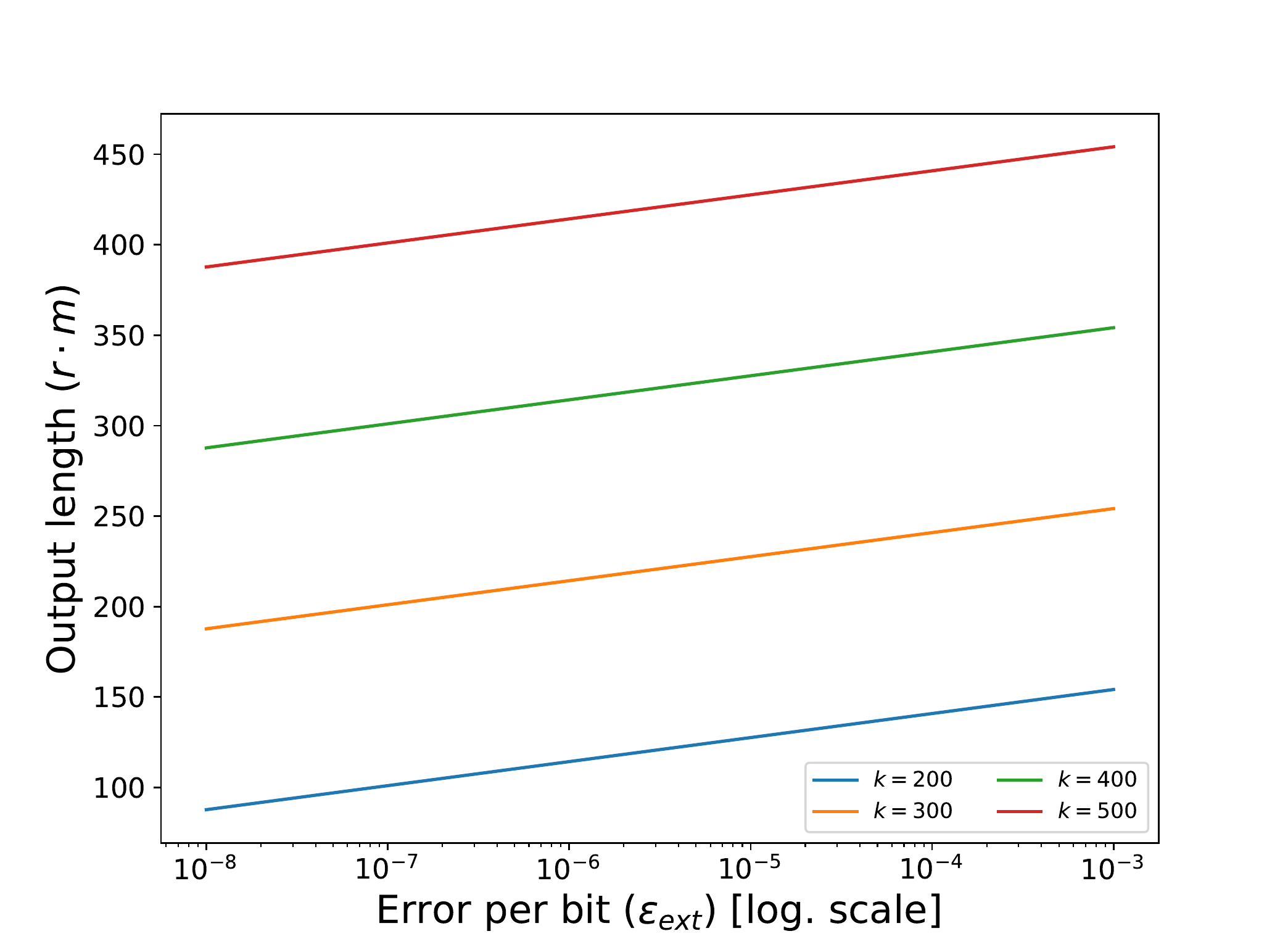}
\caption{\textbf{Ratio between the output length of the extractor and the error per bit parameter.} This figure shows linear dependence in semi-log scale of the output length $m$ (multiplied by a constant factor $r$) as a function of the error per bit $\epsilon_{ext}$. Output length increases with the error and the greater the min-entropy $k$, the greater the output length. This plot refers both to the Standard weak design \cite{Nisan} as well as the Block weak \cite{framework} design (which we adopt in our paper) algorithms.}
\label{fig:extr_3}
\end{figure}

\begin{figure}[H]
\centering
\includegraphics[scale=0.69]{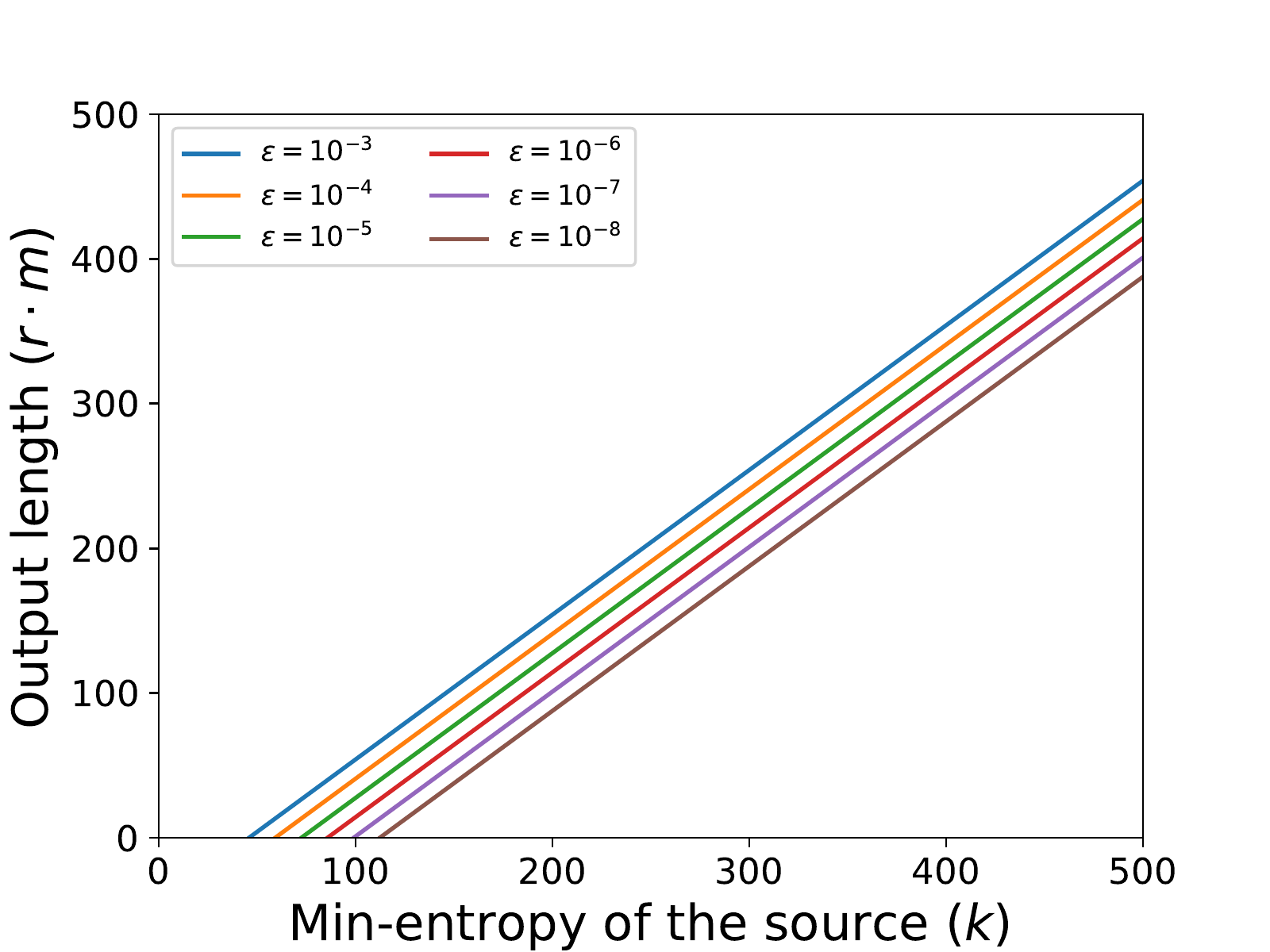}
\caption{\textbf{Final random string length in terms of the min-entropy of the source.} It can be noted that the output length $m$ (multiplied by a constant factor $r$) is a linear growing monotone function of the min-entropy of the source $k=\mathcal{H}_{min}^{\epsilon}$ and it also increases with the error per bit $\epsilon_{ext}$. This plot refers both to the Standard weak design \cite{Nisan} as well as the Block weak \cite{framework} design (which we adopt in our paper) algorithms.}
\label{fig:extr_4}
\end{figure}

\begin{figure}[H]
\centering
\includegraphics[scale=0.69]{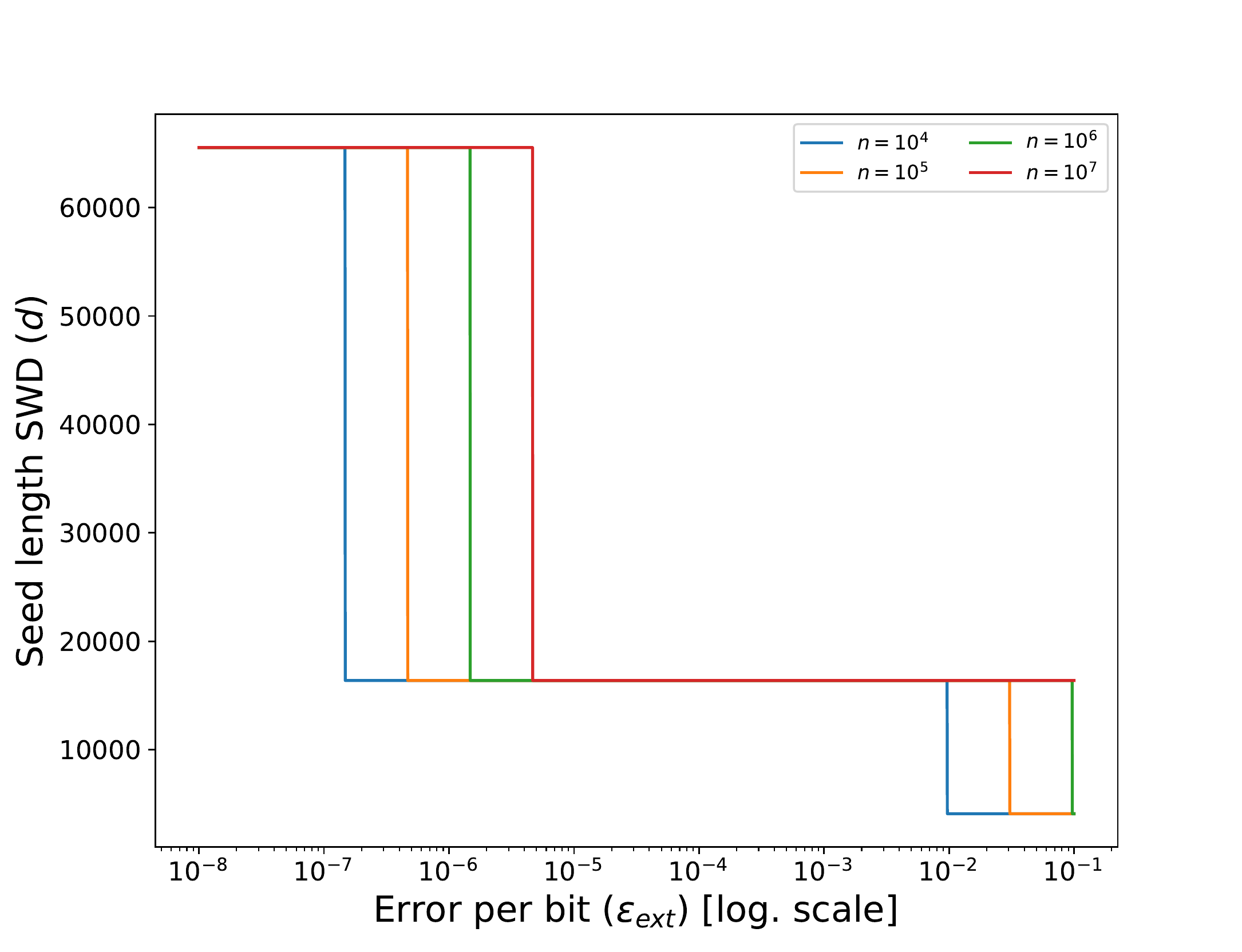}
\caption{\textbf{Dependency of seed length for the Standard weak design from the error per bit.} Here different graphs of the seed length $d$ as a function of the error per bit $\epsilon_{ext}$ are plotted, for different input source length in semi-log scale. These functions are step functions and, with the same error, the seed increases with the size of the input $n$. This plot refers to the Standard weak design algorithm \cite{Nisan}.}
\label{fig:extr_5}
\end{figure}

\begin{figure}[H]
\centering
\includegraphics[scale=0.69]{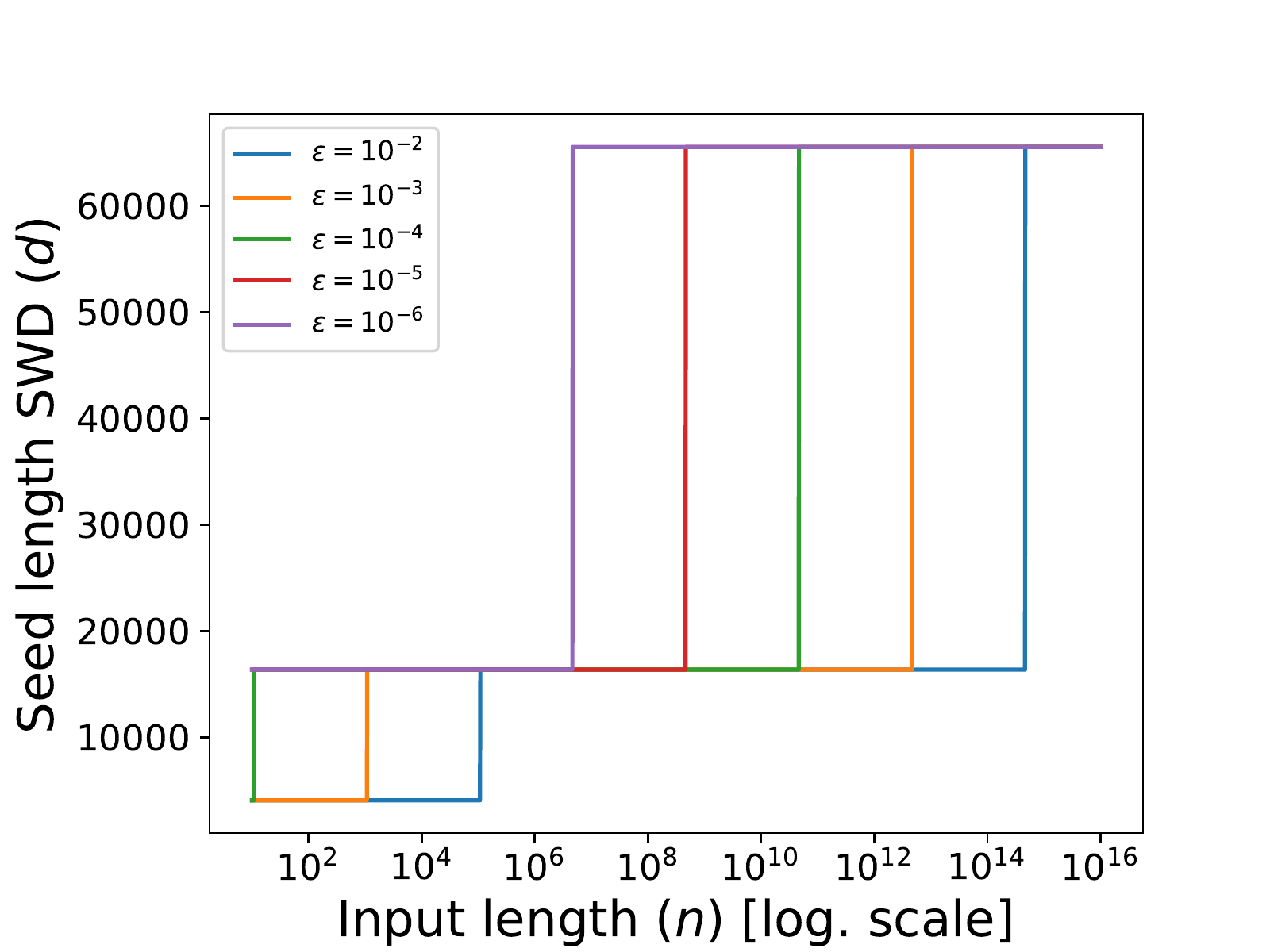}
\caption{\textbf{Comparison between seed length and the input length of the source.} In this figure is represented the seed length $d$ vs the input length $n$ in semi-log scale, plotted for different error per bit $\epsilon_{ext}$ parameters. The seed is a step function of the input length, it increases with $n$ and, with the same input length, the seed length is greater for lower errors. This plot refers to the Standard weak design algorithm \cite{Nisan}.}
\label{fig:extr_6}
\end{figure}

\begin{figure}[H]
\centering
\includegraphics[scale=0.75]{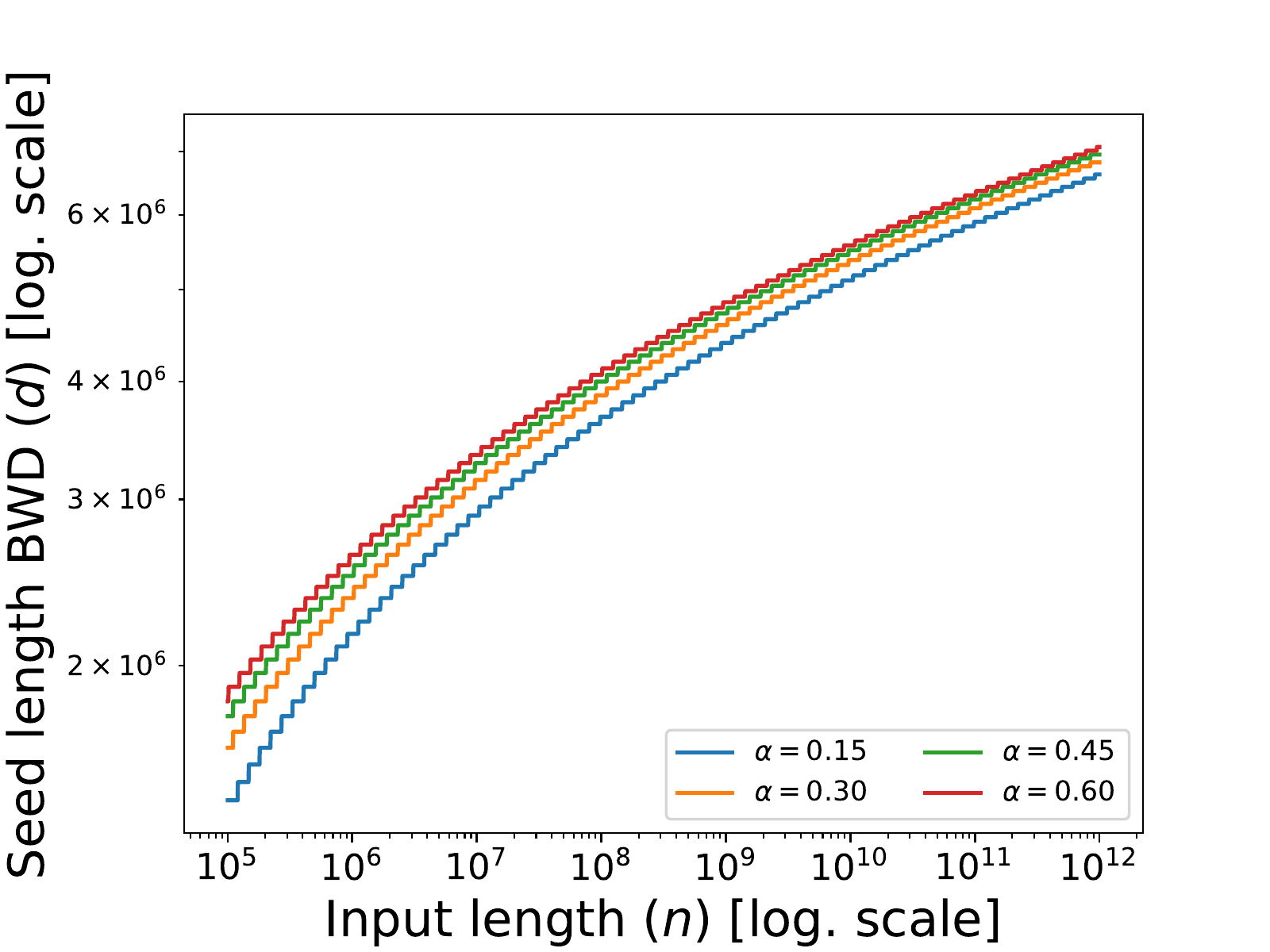}
\caption{\textbf{Relation between the seed length of Block weak design and the input length of the source.} The seed length $d$ as a function of the input length $n$ is plotted for different values of min-entropy per bit $\alpha$ and the error per bit $\epsilon_{ext}$ parameter is fixed at $\epsilon_{ext} =10^{-7}$. $d$ is a monotone increasing function of $n$ and it increases also with $\alpha$. Both the axes are in logarithmic scale.  This plot refers to the Block weak design algorithm, which was used in our paper, \cite{framework}.}
\label{fig:extr_7}
\end{figure}

\begin{figure}[H]
\centering
\includegraphics[scale=0.75]{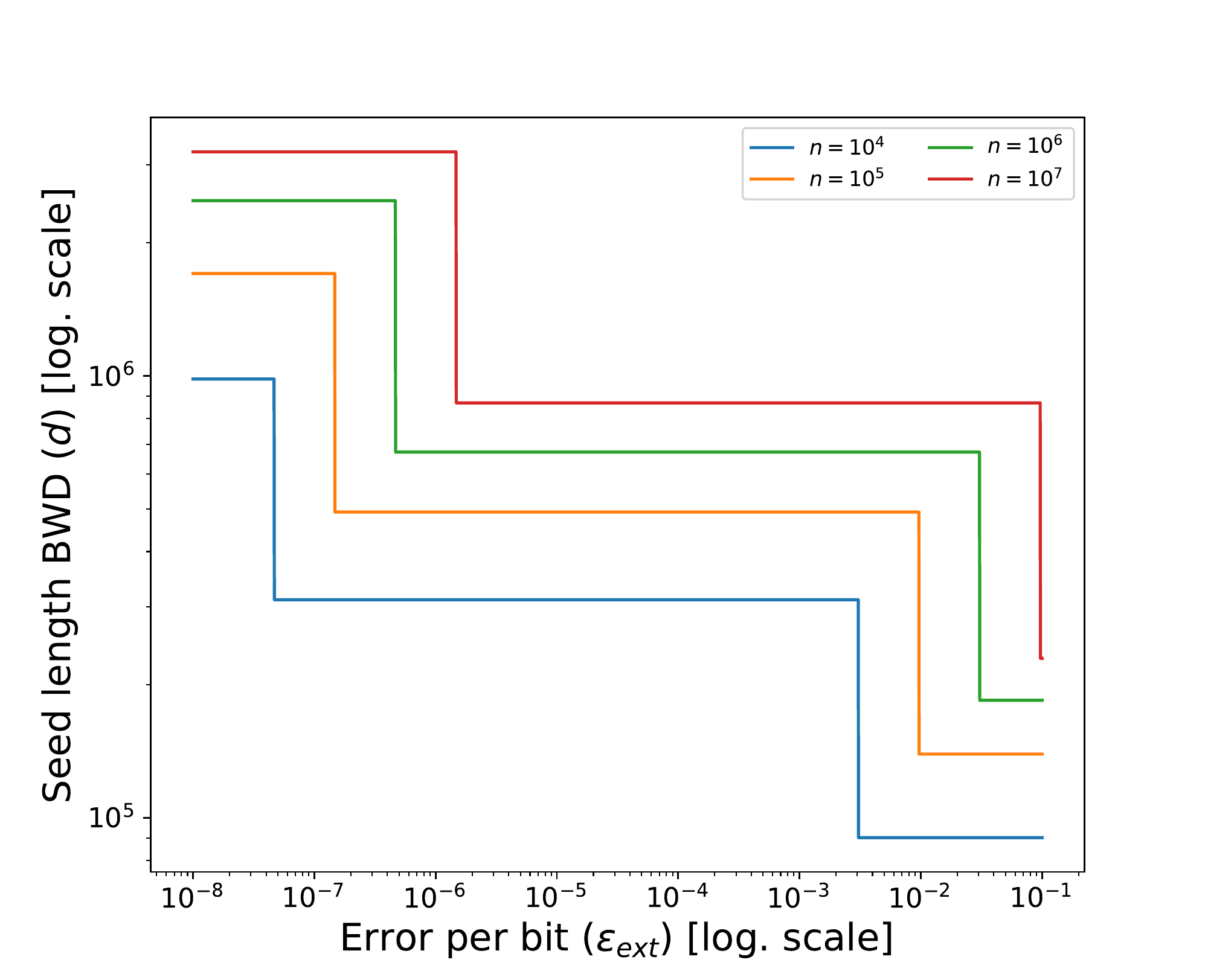}
\caption{\textbf{Seed length of Block weak design depending on the error per bit.} For different input length of the source, the seed length $d$ as function of the error per bit $\epsilon_{ext}$ is plotted. $d$ is a descending step function of $\epsilon_{ext}$, but it increases with $n$. The min-entropy is fixed at $\alpha =0.4$ and both the axes are in logarithmic scale. This plot refers to the Block weak design algorithm, which was used in our paper, \cite{framework}.}
\label{fig:extr_8}
\end{figure}

In Fig.\ref{fig:extr_1}-\ref{fig:extr_8}, we show the behaviour of our extractor's parameters.

